\documentclass[useAMS,usenatbib]{mnras}

\newcommand{\msun}{M$_\odot$}
\newcommand{\lsun}{L$_\odot$}

 \usepackage{graphicx}

\title[Formation of And II]{Formation of Andromeda II via a gas-rich major merger and an interaction with M31}

\author[S. Fouquet et al.]
{Sylvain Fouquet$^{1}$, Ewa L. {\L}okas$^{1}$, Andr\'es del Pino$^{1}$ and Ivana Ebrov{\'a}$^{1,2}$\\
$^{1}$Nicolaus Copernicus Astronomical Center, Polish Academy of Sciences, Bartycka 18, 00-716 Warsaw, Poland\\
$^{2}$Astronomical Institute, Czech Academy of Sciences, Bo\v{c}n\'{i} II 1401/1a, CZ-141 00 Prague, Czech Republic\\}

\begin{document}

\date{}

\pagerange{\pageref{firstpage}--\pageref{lastpage}} \pubyear{2016}

\maketitle

\label{firstpage}

\begin{abstract}
Andromeda II (And II) has been known for a few decades but only recently observations have
unveiled new properties of this dwarf spheroidal galaxy. The presence of two stellar populations, the bimodal star
formation history (SFH) and an unusual rotation velocity of And II put strong constrains on its formation and
evolution. Following \citet{Lokas2014}, we propose a detailed model to explain the main properties of And II involving
(1) a gas-rich major merger between two dwarf galaxies at high redshift in the field and (2) a close interaction with
M31 about 5 Gyr ago. The model is based on $N$-body/hydrodynamical simulations including gas dynamics, star formation
and feedback. One simulation is designed to reproduce the gas-rich major merger explaining the origin of stellar
populations and the SFH. Other simulations are used to study the effects of tidal forces and the ram pressure stripping
during the interaction between And II and M31. The model successfully reproduces the SFH of And II
including the properties of stellar populations, its morphology, kinematics and the lack of gas.
Further improvements to the model are possible via joint modelling of all processes and better treatment of baryonic
physics.
\end{abstract}


\section{Introduction}

The Local Group (LG) is composed of two massive spiral galaxies, the Milky Way (MW) and M31, and a smaller one, M33.
In addition to these normal galaxies, more and more dwarf galaxies are discovered \citep[see][for an updated review of
the LG dwarf galaxies]{McConnachie2012}. The new discoveries came as a result of the SDSS \citep{Willman2005,
Sakamoto2006, Zucker2006, Belokurov2006, Belokurov2007, Walsh2007}, the PAndAS survey \citep{Ibata2007, Martin2009},
the Pan-STARRS1 $3\pi$ Survey \citep{Martin2013, Laevens2015} and most recently the Dark Energy Survey 
\citep{Bechtol2015, Koposov2015}. Most of these objects are dwarf spheroidal (dSph) galaxies: spheroidal in shape, 
devoid of gas and current star formation and typically located close to the MW or M31 ($< 300$ kpc). Except for a 
few ellipticals, the remaining LG dwarf galaxies are dwarf irregulars (dIrr): gas-rich, with high star formation 
rates and located rather far from the MW and M31 ($> 300$ kpc), with a notable exception of the Magellanic Clouds.

In the context of the $\Lambda$CDM paradigm, dwarf galaxies are understood as remnants of primordial galaxies, full of
dark matter, which have not yet merged with other objects to build massive galaxies \citep{Dekel1986}. Indeed, the LG
dIrrs have gas fractions larger than 50\% \citep{McConnachie2012} similarly to the high-$z$ galaxies
\citep{Rodrigues2012}. In the case of dSphs, their properties (the lack of gas and the spheroidal shape) are explained
by a long-term interaction with their host galaxies due to tidal forces, dynamical
friction and ram pressure stripping \citep{Mayer2001, Mayer2007, Kliment2009, Kazan2011a, Lokas2011, Lokas2012,
Nichols2015}.

In this study, we investigate specifically the origin of the Andromeda II (And II) dSph galaxy. This dSph discovered in
1972 \citep{VdB1972} lies in the vicinity of M31 at a distance of 184 kpc \citep{McConnachie2012}. It seems to be a
typical, relatively luminous dSph devoid of gas, without star formation for the last several Gyr \citep[][Hidalgo et al.
in preparation]{Weisz2014, Skillman2016}, with a low central surface brightness of 24.8 mag arcsec$^{-2}$ and a
quasi-spheroidal shape with ellipticity close to 0.2 \citep[][del Pino et al. in preparation]{McConnachie2007}.
However, when studied in more detail, And II reveals a complicated star formation history (SFH) with two star formation
episodes (Hidalgo et al. in preparation) which generated two stellar populations: one more extended containing
primordial stars ($>$ 11 Gyr old) and one more spatially concentrated composed of intermediate-age stars (5-9 Gyr old)
\citep[][del Pino et al. in preparation]{McConnachie2007}. In addition, \citet{Ho2012} measured a clear rotation signal
in And II, which is unusual for a dSph in the LG. Moreover, this rotation is around the major axis of the dwarf and not
the minor one as commonly observed in spiral galaxies. \citet{Amorisco2014} reanalyzed the same kinematic data and
claimed detection of substructure that could form a stellar stream. They speculated that such substructure could result
from a merger between And II and a smaller object.

Until recently, mergers between dwarf galaxies in environments such as LG were believed to be rare events due to large
relative velocities of the possible progenitors. However, investigations of subhalo statistics in simulated LG-like
environments revealed \citep{Klimentowski2010, Deason2014} that a non-negligible part (10-30\%) of present-day dwarf
galaxies could have undergone a major merger mainly at high redshift when they were not yet bound to their hosts
and when the galaxy density in the Universe was higher. These mergers turned out to take place between
galaxies accreted by the host in pairs, which explains their low relative velocities. \citet{Kazan2011b} resimulated
some of such pair-like configurations, populating the subhaloes with stellar disks and demonstrated that such mergers
can lead to the formation of dSph-like objects. Some dwarf galaxies
of the LG could therefore be remnants of major mergers which have occurred in the field a few Gyr ago. In a recent study
\citet{Benitez2016} investigated major mergers between gas-rich dwarf galaxies in a constrained simulation of the LG.
They showed that products of such mergers would indeed have two stellar populations with different spatial
distributions similar to the ones seen in And II.

\citet{Lokas2014} proposed the first detailed model for the origin of And II as the result of a major
merger of two disky dwarf galaxies of equal mass. This study used collisionless $N$-body simulations with the
progenitor galaxies composed of a stellar disk and a dark matter halo. The different spatial distribution of the two
stellar populations was explained by the different size of the stellar disk in the two progenitors.
The unusual prolate rotation was explained as the result of a head-on collision with a particular
orientation of the progenitor disks. \citet{Ebrova2015} generalized this result by considering other initial
configurations in terms of the orbit and inclination of the disks. They demonstrated that prolate rotation can result
from a variety of initial conditions and does not require fine tuning thus adding credibility to the model. They
have also shown that it is practically impossible to produce such type of rotation via tidal stirring of initially
disky dwarfs orbiting a MW-like host \citep[e.g.][]{Lokas2015}. Although tidal stirring can be very efficient in
replacing rotation by random motions of the stars, it cannot induce rotation around the major axis
and whatever rotation remains in the dwarf is always around the minor axis.

These studies did not however include gas dynamics and star formation so could not
address the origin of stellar populations of different age. If the major merger occurred at high redshift in the field,
it is highly probable that the two dwarf galaxies were gas-rich.
In this work we therefore extend the model proposed by \citet{Lokas2014} by adding gas and star formation to the
picture. We present a complete model for And II since the beginning of the gas-rich major merger (10 Gyr ago) till the
present day. In order to be complete,
this study also takes into account the interaction between M31 and the remnant of the gas-rich major merger in order to
strip the gas. In sections 2 and 3, we describe the gas-rich major merger: the initial conditions, the numerical
configuration and the final remnant. Then in sections 4 and 5, we investigate the interaction with M31
taking into account the possible orbit of And II around M31 and the effects of the tidal force and the ram pressure
stripping. Finally, sections 6 and 7 present the discussion of our results and conclusions from this work.

\section{Initial parameters of the major merger}

In our scenario, And II is the result of a major merger which occurred at
high redshift ($z > 2-3$) in the field between two dwarf galaxies. Due to the lack of dwarf galaxy observations
at high redshifts, it is uncertain what the progenitor galaxies of And II could look like. We therefore have to make
some assumptions to define the main properties of And II progenitors using what is thought reliable concerning high
redshift dwarf galaxies. First, we assume that the two progenitors are identical. This choice leads to a 1:1 merger
and decreases the number of parameters by a factor of two (see Table \ref{tab:prop_prog}).
It also seems essential to reproduce the And II kinematics following \citet{Lokas2014} and \citet{Ebrova2015}.
Consequently, we did not try to change the mass ratio. However, one may argue that a major merger is mandatory in
order to produce a dSph galaxy from two disky ones. A minor merger (with a mass ratio below $1/3$) would only perturb
the biggest galaxy without significantly affecting its morphology and kinematics \citep[see e.g.][]{Kazan2011b}.
Second, as the merger occurs at high redshift in the field, we expect the two dwarf galaxies to be dIrrs, thus they are
gas-rich and disky.

In this section, we present and explain the initial conditions of the gas-rich major merger. The final set of
parameters used in our simulation was selected in order to provide best match to the observations and was
determined as a result of many numerical experiments we performed with different combinations of their values.

\subsection{The stellar component}

We assumed the dwarf galaxy progenitors have a stellar disk with an exponential density profile defined by three
most important parameters: the mass, the radial extension and the thickness. And II has a $V$-band
absolute magnitude of $-12.6$ mag \citep{Kalirai2010}. By assuming a mass-to-light ratio of 1 \msun/\lsun\ as in
\citet{Kalirai2010}, the total stellar mass is estimated at $9.1 \times 10^6$ \msun. However, as And II's stellar
population is mainly composed of primordial and intermediate-age stars \citep[][Hidalgo et al. in
preparation]{Weisz2014}, the present mass-to-light ratio could be larger than 1. This implies that the total stellar
mass of the two progenitors should be larger than $9.1 \times 10^6$ \msun. In our simulation, we have chosen the stellar
mass per progenitor of $9 \times 10^6$ \msun, so the total stellar mass of $18 \times 10^6$ \msun. Different
test simulations made clear that this large initial mass (about twice the final one) is needed because some of the stars 
are ejected from the remnant or are too faint to be visible.

The second important parameter is the radial size of the disk. As the progenitor disks have an exponential shape, it is
defined by the scale radius, $R_{\rm d}$, which should be between 0.4 and 1 kpc. Values smaller than 0.4 would lead to
the formation of a remnant which is too dense, larger ones would make it too extended. We have tested different
scale radii between 0.4 and 1.0 kpc. Finally we have chosen a scale radius of 0.8 kpc. This parameter, as
expected, turned out to be the main parameter that needed to be adjusted in order to accurately reproduce the slope of
the stellar density profile. The adopted thickness of the disk, defined by $z_{\rm d}/R_{\rm d} = 0.25$, is a typical
value for dwarf galaxies. We did not vary its value in the test simulations, as it seems to have little
effect on the final outcome of the merger.

\begin{figure}
  \begin{center}
    \centering
    \includegraphics[width=1\linewidth]{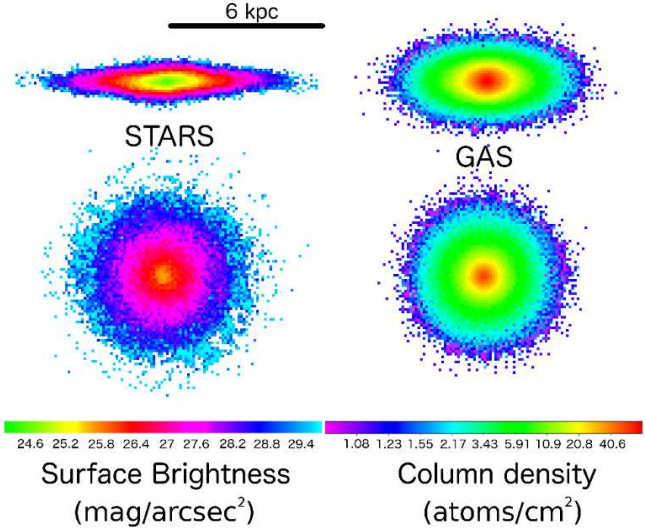}
    \caption{Images of the stars (left) and gas (right) distribution in the progenitors. The thick line
indicates the scale length of the images. For the stars, colours code the surface brightness (in mag arcsec$^{-2}$)
with an assumed mass-to-light ratio of 1 \msun/\lsun. For the gas, colours code the column density of the gas (in $10^{19}$
atoms cm$^{-2}$). Top panels show the two components viewed edge-on while in the bottom panels they are viewed face-on.}
\label{fig:ic_dwarf}
  \end{center}
\end{figure}

\subsection{The gas component}

The dIrrs in the LG \citep{McConnachie2012} or in surveys like LITTLE THINGS \citep{Hunter2012} have high gas fractions
which can reach more than 80\% of the baryonic mass. At high redshift, this value must have been statistically larger
because less gas was consumed by star formation. Observations of And II do not provide any hint to derive the gas
fraction in the past because it seems to have no more gas at present \citep{Grcevich2009}. The gas fraction
was chosen to be 70\% of the baryonic mass. With the stellar mass of $9 \times 10^6$ \msun, this leads to the gas mass
of $21 \times 10^6$ \msun\, and the total baryonic mass of $30 \times 10^6$ \msun. In our test simulations, we
have also tried other gas fractions from 60\% to 80\%. As expected, an increase (decrease) of the gas fraction increases 
(decreases) the star formation. Nevertheless, the star formation also depends on feedback, therefore we have decided to keep
the gas fraction constant at 70\% and to fine tune the feedback. This obviously shows that the problem is degenerated as
two different sets of parameters can provide the same result. This is not surprising due to the large number of
parameters and the limited and uncertain observations.

We also always assumed that the gas disk has the same extension as the stellar disk (see Figure \ref{fig:ic_dwarf}). 
Again, the gas extension has a direct effect on the star formation; increasing it will decrease the star formation because 
the gas density will decrease. On the other hand, the thickness of the gaseous disk is not a free parameter because it depends 
on the balance between the pressure force and the gravitational force in this component. Consequently, the gas disk we obtain 
is about a factor of 1.5 thicker than the stellar disk (see Figure \ref{fig:ic_dwarf}).

\subsection{The dark matter component}

The stellar and gas disks are embedded in a dark matter halo. Different works \citep[][and reference therein]{Kirby2014} 
tend to show that the dark mass-baryonic mass ratio can be very large, even larger than a hundred for faint dwarf galaxies 
like Sextans. We have chosen the ratio of the dark to baryonic mass to be equal to 33. This choice is mainly motivated by 
the need to reproduce the level of velocity and velocity dispersion in the And II kinematics. As the initial baryonic mass 
in the simulation is constrained by the actual And II's stellar mass, a larger (smaller) mass-to-light ratio would imply 
a larger (smaller) dark matter mass and thus larger (smaller) kinematics. In our test simulations we tried mass-to-light 
ratios between 20 and 50.

The density profile of dark matter is assumed to be similar to the Navarro-Frenk-White (NFW) profile
\citep{NFW1996} with a cut-off at the virial radius and the total mass equal to $10^9$ \msun $\,$ but with a shallower
inner slope of $-0.6$ instead of $-1$. This choice was made to better match the inner kinematic profile. With a
pure NFW profile, dark matter mass is too concentrated and the inner slope of the rotational velocity is too steep. We
also note that this shallower slope is consistent with recent findings concerning the effect of baryons on dark matter
distribution \citep{Gover2010, Gover2012}. In our case this flattening of the dark matter slope would have been caused
by the first generation of stars prior to the epoch we attempt to model.

\begin{table}
  \caption{Properties of different components of the progenitors of the gas-rich major merger.}
  \begin{center}
    \begin{tabular}{llr}
      \hline \hline \\
      Components  & Properties                    & Values                 \\ \hline
      Dark Matter & total  mass                   & $         10^9$ \msun  \\
                  & inner slope                   & $-0.6$                 \\
                  & outer slope                   & $-3.0$                 \\ \hline
      Baryons     & baryonic mass                 & $30 \times 10^6$ \msun \\
                  & baryon fraction               & 0.03                   \\ \hline
      Stars       & stellar mass                  & $9 \times 10^6$ \msun  \\
                  & disk scale-length $R_{\rm d}$ & 0.8 kpc                \\
                  & disk thickness                & 0.2 kpc                \\ \hline
      Gas         & gas mass                      & $21 \times 10^6$ \msun \\
                  & gas fraction                  & 70\%                   \\
                  & disk scale-length $R_{\rm d}$ & 0.8 kpc                \\ \hline
    \end{tabular}
    \label{tab:prop_prog}
  \end{center}
\end{table}

\subsection{Creating the progenitors}

To create the initial conditions for the progenitors, we used the method described in \cite{Widrow2005} and
\cite{Widrow2008}. It allows to create stable numerical realizations of galaxies in near-equilibrium, composed of a
halo and a stellar disk. This method produces objects closer to equilibrium and therefore gives us better control over
the structural parameters compared to the one described in \cite{Hernquist1993} because it is based on the distribution
functions, not just their moments. However, this method does not allow to define the initial conditions for the gas. In
order to include the gas component we therefore split the baryonic disk particles randomly into two populations: gas
particles and stellar particles matching the assumed gas fraction (70\%). We end up with a dwarf galaxy with properties
given in Table~\ref{tab:prop_prog} but unstable due to the gas.

\begin{figure}
  \begin{center}
    \centering
    \includegraphics[width=0.9\linewidth]{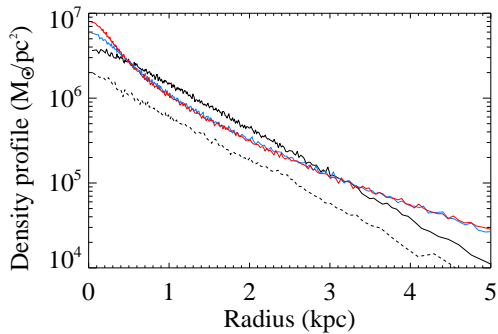}
    \caption{Gas density profiles at the beginning of the initial condition simulation (black line), at the end after
10 Gyr (blue line) and after 0.5 Gyr more when the star formation and feedback are on (red line). The dashed black
line represents the density profile of stars which remains the same during the steps of the formation of the
progenitors.}
\label{fig:denspro_gas}
  \end{center}
\end{figure}

\begin{figure}
  \begin{center}
    \centering
    \includegraphics[width=1\linewidth]{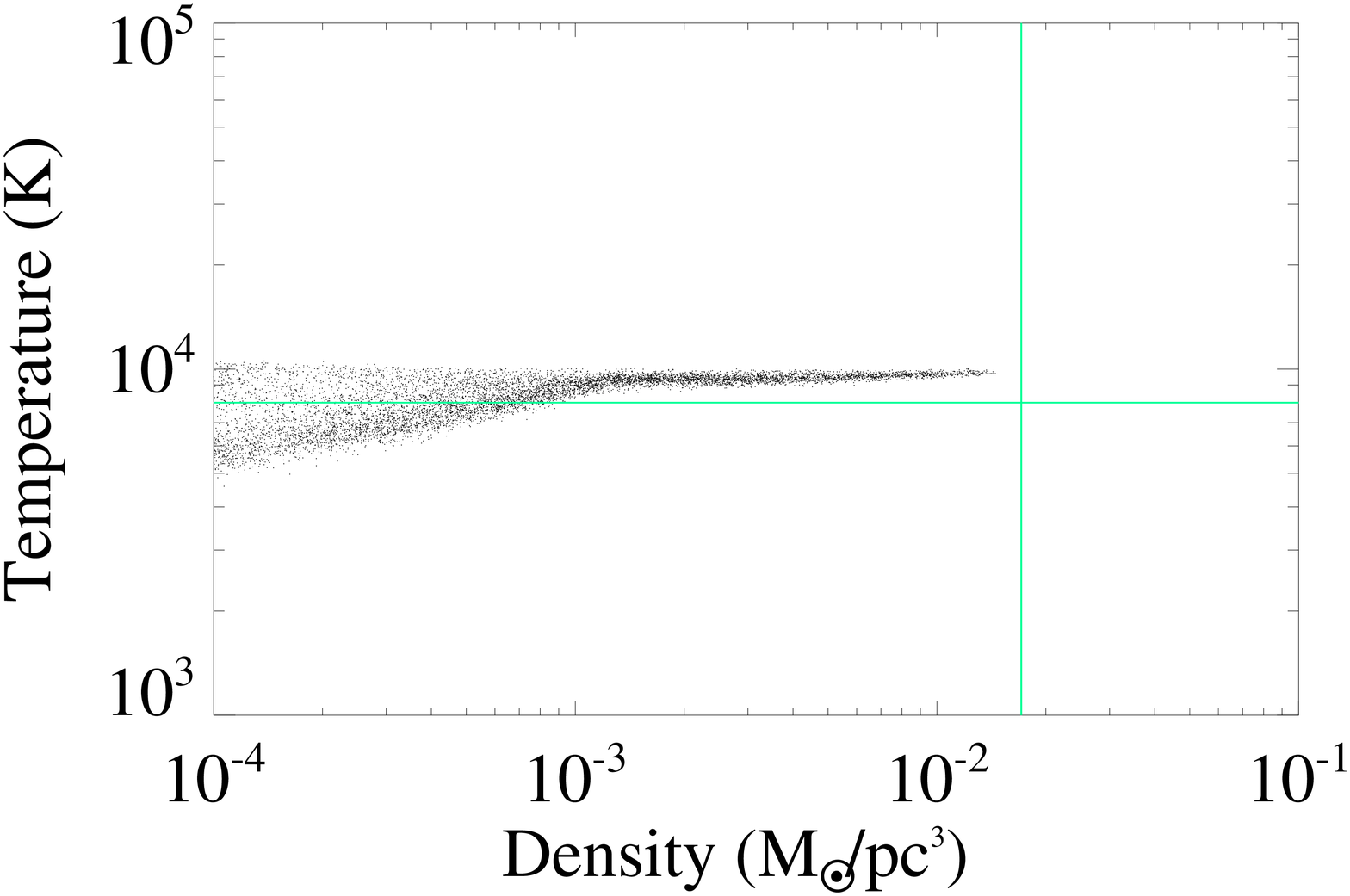}
    \caption{Gas density-temperature plot of the stable progenitors after 0.5 Gyr of evolution with the star formation
on. The vertical line shows the density threshold for the star formation. The horizontal one indicates the initial
temperature of all gas particles: $8000$ K.}
\label{fig:temp_dens_gas}
  \end{center}
\end{figure}

In order to avoid gas instabilities due to the gas pressure, we let the dwarf galaxy evolve during 10 Gyr by setting
the initial gas temperature to $8000$ K. For this simulation we used the publicly available version of the GADGET2
$N$-body/SPH code \citep{Springel2001, Springel2005}. During this simulation, there is no star formation because our
purpose here is just to stabilize the gas component rather than to investigate the evolution of the galaxy in
isolation. After 10 Gyr, the dwarf galaxy is stable again.

During this evolution the stellar disk becomes thicker (by less than $10 \%$) but has the same exponential shape and
the same kinematics. The gas is almost 1.5 times thicker than the stars, its radial extension remains similar to the
stars, and the gas loss is small ($<$ 10\%). Stars are all conserved so the new gas fraction is 67\%, close to the
previous one, 70\%. The evolution also results in a slight overdensity of gas close to the centre (see Figure
\ref{fig:denspro_gas}). However, there is no need to get a perfect exponential shape for the gas because when the star
formation with feedback is turned on, it affects the gas density profile, increases the gas bump at the centre and
keeps the maximal temperature at $10\,000$ K due to the cooling (see Figure \ref{fig:temp_dens_gas}). A much lower
initial temperature, for example 20 K instead of 8000 K, allows us to reduce the pressure force leading to less gas
escaping in the external parts. However, it creates a much larger gas overdensity in the centre.

\begin{figure}
  \begin{center}
    \centering
    \includegraphics[width=1\linewidth]{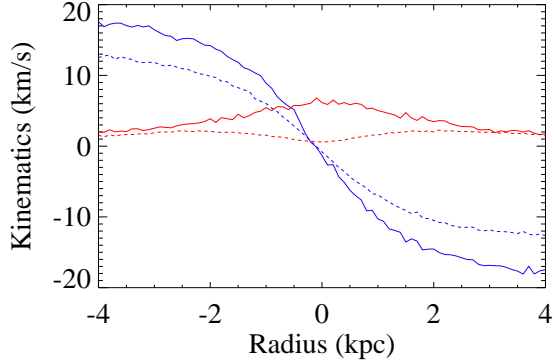} \\
    \caption{The kinematics (rotation velocity in blue, velocity dispersion in red) of the gas (dashed lines)
and stars (solid lines) after 10 Gyr in the initial conditions simulation. Stars and gas have similar
rotation although the gas has a smaller rotation level (at 2 kpc, the absolute difference is 4.6 km s$^{-1}$ and a 
relative one about 30\%). The stellar velocity dispersion has a maximum at the centre ($\sim 7$ km s$^{-1}$) to support 
the stellar component. On the contrary, the gas velocity dispersion is smaller with a drop at the centre because it is 
the pressure force and not the velocity dispersion that stabilizes the gas component.}
\label{fig:kine_gas}
  \end{center}
\end{figure}

The rotating gas kinematics is similar to that of the stars although the rotation velocity is smaller due to the
initial instability. However, gas rotation is not important in our study because the gas is stripped in the end.
The small fraction that turns into stars does not significantly affect the overall kinematics of the final remnant.
The velocity dispersion of the gas is lower than for the stars, almost constant with radius with a slight drop at the
centre. This property is due to the gas pressure which does not need velocity dispersion to keep the gas stable (see
Figure \ref{fig:kine_gas}).

\begin{table}
  \caption{Numerical properties of the gas-rich major merger.}
  \begin{center}
    \begin{tabular}{lrrr}
      \hline \hline \\
      Properties            & Dark matter & Stars  & Gas     \\ \hline
      Number of particles   & 200 000     & 60 000 & 140 000 \\
      Particle mass (\msun) & 5114.6      & 149    & 149     \\
      Softening length (pc) & 60          & 20     & 20      \\ \hline
    \end{tabular}
    \label{tab:param_simu}
  \end{center}
\end{table}

\subsection{The orbit of the major merger}

\begin{figure}
  \begin{center}
    \centering
    \includegraphics[width=1\linewidth]{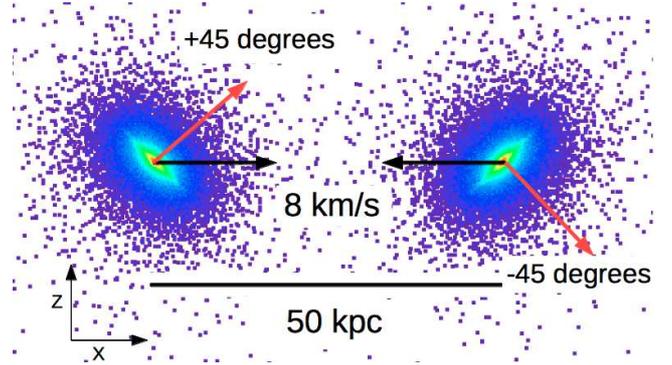}
    \caption{Initial conditions of the major merger viewed in the $xz$ plane of the simulation box.
The dwarf images include both stars and gas. The red arrows
indicate the direction of the disk angular momenta which lie in the $xz$ plane. The velocities and the centres of the
dwarf galaxies are aligned with the $x$-axis.}
\label{fig:IC_orbit}
  \end{center}
\end{figure}

A major merger between galaxies can result from a variety of orbital initial conditions. \citet{Khochfar2006}
studied the statistics of orbits of merging massive haloes in a cosmological simulation and found that an eccentricity
around one, corresponding to a parabolic orbit, is the most likely. Unfortunately, no such statistics is available
for smaller mass haloes or dwarf galaxies in LG-like environments. However, \citet{Ebrova2015} demonstrated that
for a major merger with a hyperbolic orbit and a small pericentre after the first passage the orbital angular
momentum is lost and the trajectory evolves from a hyperbola to a highly eccentric ellipse for the second passage,
which is similar to an almost radial orbit. A similar result is expected with a parabola which is a limiting case, $e =
1$, of a hyperbola. In our model there are also two passages. The first one could be on a parabolic trajectory and a
close pericentre which would lead to a nearly head-on second passage. In order to simplify the initial conditions, we
start with a head-on collision. Therefore at $t = 0$, the two dwarf galaxies are separated by 50 kpc and they are
approaching with a velocity of 8 km s$^{-1}$ along the $x$-axis of the simulation box, which corresponds to the initial
relative velocity of 16 km s$^{-1}$ (see Figure \ref{fig:IC_orbit}).

The disks of the progenitors are rotated so that the disk angular momenta have the same component along the $x$-axis.
This implies that $L_{x,1} = L_{x,2}$, $L_{y,1}=-L_{y,2}$ and $L_{z,1}=L_{z,2} = 0$. This leaves us only the freedom
to choose the inclination of the angular momentum between 0 and 90 deg for the first galaxy (the left one in Figure
\ref{fig:IC_orbit}). With an angle of 0 deg, the total angular momentum is along
the positive $x$. With 90 deg, the angular momentum is perpendicular to the $x$-axis. A value of 0 deg for the
first progenitor implies 0 deg for the second too. The dependence of the remnant properties on this parameter has been
also studied by \citet{Ebrova2015}. Following this study, we can expect that small inclination angles would provide
a large angular momentum for the remnant galaxy but not a spheroidal shape. On the contrary, a choice of a high
inclination angle (close to 90 degrees) would produce a galaxy without significant prolate
rotation. We have therefore chosen an angle of 45 deg, as in \citet{Lokas2014}, for the first galaxy, and $360 - 45
= 315$ deg =($-45$ deg) for the second one to match observations (see Figure \ref{fig:IC_orbit}).

\subsection{Numerical prescription for the star formation}

For the simulations described below we used a modified version of the GADGET2 $N$-body/SPH code \citep{Springel2001,
Springel2005} with star formation, feedback and cooling processes added as described in \citet{Hammer2010} and 
\citet{Wang2012} following recipes defined in \cite{Cox2006}. Different parameters are available to tune the star 
formation. We choose to mainly follow \cite{Cox2006}, i.e. we adopt the star formation efficiency of 0.03, the feedback 
index of 2 and the density threshold for star formation of 0.0171 \msun pc$^{-3}$. Our assumptions differ only in one 
feedback parameter, $\tau_{\rm fb}$, determining the time-scale of feedback thermalization. When $\tau_{\rm fb}$ is large 
the energy due to the star formation is quickly released and vice versa. This parameter is tuned in order to get the star 
formation rate which fits the observed SFH in And II. The value in the presented simulation is 5.5 Myr which equals to 0.7 
times the normal feedback timescale defined by \cite{Cox2006}. The numbers of particles, the softening lengths and the 
masses of the particles are provided in Table \ref{tab:param_simu}.

\section{Results of the major merger}

\begin{figure}
  \begin{center}
    \centering
    \includegraphics[width=0.9\linewidth]{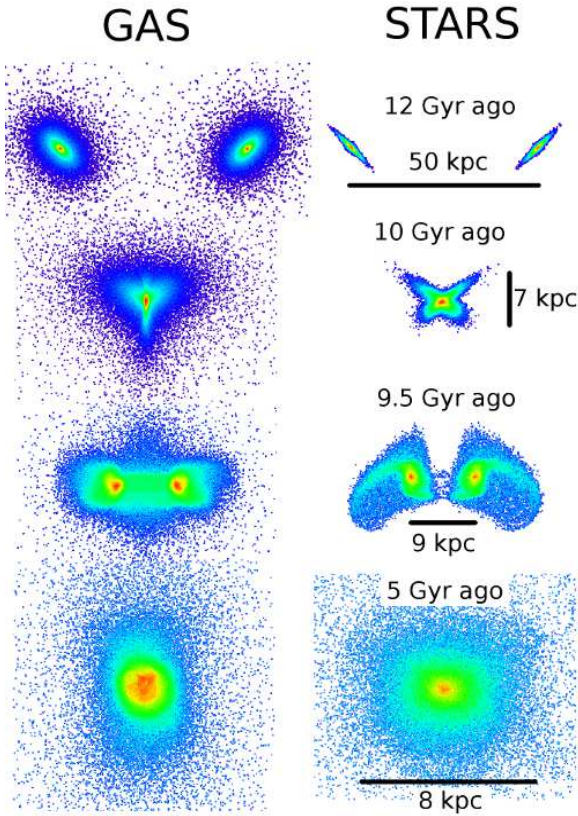}
    \caption{Main phases of the gas-rich major merger for the gas (left panels) and the stars (right panels). The view
is always in the $xz$ plane of the simulation box as in Figure \ref{fig:IC_orbit}. The rows of panels from top to
bottom show, respectively, the initial conditions 12 Gyr ago, the first passage 10 Gyr ago, the apocentre 9.5 Gyr ago
and the end of the star formation 5 Gyr ago. Times and scales are indicated for each phase. The scales are the same
for stars and gas but change with the phase. The colours code the stellar and gas mass density in logarithmic scale.
The colour coding is different for each phase in order to clearly show the density gradient.}
\label{fig:ic_merger}
  \end{center}
\end{figure}

\subsection{Star formation history}

The simulation starts with two separated dwarf galaxies. The stars inside these galaxies were formed during the
first Gyrs of the evolution of the Universe. We did not try to reproduce this initial star
formation period due to its complexity which must have involved gas accretion, formation of primordial stars,
cosmological evolution, etc. We only know that stars older than 11 Gyr represent about 70\% of all stars as seen in
observations. We assume that the simulated SFH matches the observed SFH during this early period. Our
simulation becomes consistent with the data around 11 Gyr ago.

\begin{figure}
  \begin{center}
    \centering
    \includegraphics[width=1\linewidth]{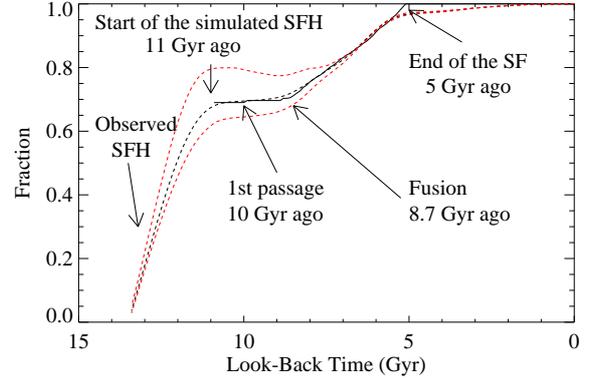}
    \caption{Simulated SFH of And II (black solid line) compared to the observed one (black dashed line) at the centre
of And II within the radius of 0.52 kpc (Hidalgo et al., in preparation). The red dashed
lines indicate the observed SFH uncertainties. The main stages of the major merger are indicated.}
\label{fig:res_majmer_SFH}
  \end{center}
\end{figure}

Between 11 and 10 Gyr ago, there is no star formation in the dwarf galaxies, which are then approaching each other, 
because the gas is not dense enough (see Figure \ref{fig:temp_dens_gas}). Ten Gyr ago, the two progenitors collide 
for the first time (the first passage). Due to the high relative velocity of the two dwarf galaxies ($\sim$ 60 km s$^{-1}$), 
the collision is quick and the stellar disks are not completely destroyed. They are warped and tidal tails form for each 
dwarf galaxy (see Figure \ref{fig:ic_merger}). The situation is different for the gas. In addition to the gravitational 
force, the gas dynamics is also influenced by pressure forces. Consequently, the gas does not behave collisionlessly. 
When the dwarf galaxies collide, the gas is compressed and its density suddenly increases. This results in the first peak of
star formation at this time. However, this star formation episode lasts only for a very short time, less than 0.1 Gyr,
because the dwarf galaxies move quickly away from each other reducing the gas density. Thus, this peak of star
formation rate produces very few new stars and is barely visible in the SFH (see Figure \ref{fig:res_majmer_SFH}).

Between 10 Gyr and 8.7 Gyr ago, the dwarf galaxies are separated again reaching their apocentre 9.5 Gyr ago (see Figure
\ref{fig:ic_merger}). Afterwards, they fall towards each other for the second passage 9 Gyr ago. It is quickly followed
by a complete fusion of the galaxies which happens 0.3 Gyr later, 8.7 Gyr ago. After that, the centres of the dwarf
galaxies are indistinguishable (see Figure \ref{fig:ic_merger}). The fusion has a clear effect on the SFH with a
strong and constant enhancement of star formation (see Figure \ref{fig:res_majmer_SFH}). The SFH slope becomes steep
because the star formation rate is now steady ($4 \times 10^{-4}$ \msun yr$^{-1}$). We assume that the star formation
stops $\sim$ 5 Gyr ago as required by the observations (see Figure \ref{fig:res_majmer_SFH}). The reason for this stop
is assumed to be the interaction with M31 discussed in sections 4 and 5 of this paper.

\subsection{Stellar surface density map}

\begin{figure}
  \begin{center}
    \centering
    \includegraphics[width=0.8\linewidth]{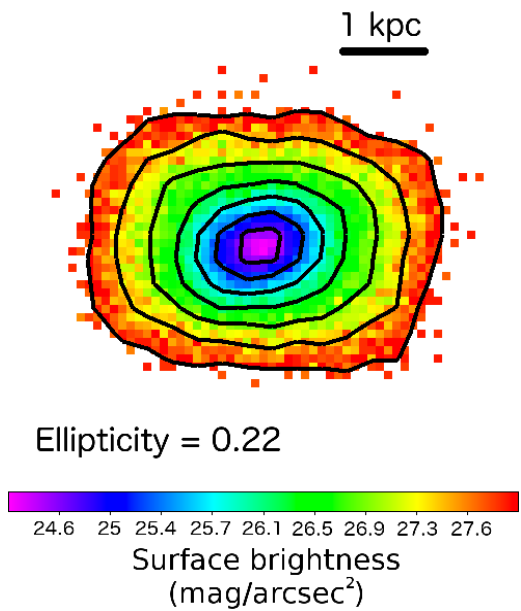}
    \caption{Image of the stellar component of the simulated merger remnant in the $xy$ plane 5 Gyr ago
obtained with a cut at 28 mag arcsec$^{-2}$. The black lines show the isophotes. The colour bar codes the surface 
brightness scale in mag arcsec$^{-2}$. The position angles of the ellipses vary a little with radius due to perturbations 
resulting from star formation.}
\label{fig:res_majmer_morpho}
  \end{center}
\end{figure}

Nearly 5 Gyr ago, when the star formation seems to end, we stop the simulation and make an image of the stellar
component which we then use to measure the luminosity and ellipticity of the remnant dwarf galaxy. For that, we have
chosen the line of sight to be the $z$-axis of the simulation box so the positions of the stellar particles are
projected onto the $xy$ plane. In fact, all lines of sight perpendicular to the $x$-axis are nearly equivalent because
the dwarf remnant is symmetrical with respect to the $x$-axis due to the initial conditions (see Figure
\ref{fig:ic_merger}). The limiting surface brightness is set to 28 mag arcsec$^{-2}$. The visible mass is converted into
the $V$-band luminosity and surface brightness, by assuming a constant mass-to-light ratio (see Figure
\ref{fig:res_majmer_morpho}). The value of the mass-to-light ratio is chosen so that the simulated $V$-band
luminosity equals to the observed one: $L_V = 9.1 \times 10^6$ \lsun. We have measured the ellipticity of the
simulated dwarf galaxy using the method described in \citet{dPM2015} by fitting iso-density contours by ellipses
(see Figure \ref{fig:res_majmer_morpho}).

The measured stellar mass for the simulated And II is $12 \times 10^6$ \msun$\,$ with a mass-to-light ratio of 1.25
\msun/$L_{\odot}$. Concerning the ellipticity, we have found $\epsilon \sim 0.22$ at the half-light radius of 0.8 kpc, 
in the $xy$ plane (see Figure \ref{fig:res_majmer_morpho}), which is within $1 \sigma$ of the observed value $\epsilon 
= 0.2 \pm 0.08$ \citep{McConnachie2006}.

\subsection{Stellar density profiles}

\subsubsection{Stellar density profiles observations}

We have computed the stellar density profiles of the two stellar populations present in And II. We used
the SFH of the galaxy (Hidalgo et al., in preparation) to assign stellar ages to the objects in the
photometric catalog obtained by \citet{McConnachie2007}. The procedure is explained in detail in \citet{dPM2015} and 
del Pino et al. (in preparation). A short summary is given here.

A synthetic color-magnitude diagram (CMD) is computed based on the SFH of And II. Using an adaptive grid,
we sample both CMDs, the synthetic and the observed one, assigning ages and metallicities to the observed stars on the
basis of their synthetic counterpart. After the sampling, two distinct stellar populations were recovered: stars older
than 11 Gyr and stars with ages between 9 and 5 Gyr. We expect this method to provide an improved age resolution
compared to the one used in \citet{McConnachie2007}. Nevertheless, even using the SFH we are not able to completely
break the age-metallicity degeneracy present in the CMD of And II. It is therefore plausible that a non-negligible
mixing between both stellar populations is still present. This seems to be the case when comparing both radial profiles
at large radii (see Figure \ref{fig:res_majmer_denspro}).

After the age determination, the radial density profiles of the two stellar populations were obtained by
fitting 38 ellipses to the isopleths of their surface density maps. We count the stars lying within each of these
regions, adopting as a galactocentric radius the average of the semi-major axes of the two boundary ellipses
delimiting a given region. Errors in the galactocentric distance come from the ellipse fitting to the isopleths, while
we assume Poissonian error for counting stars.

\subsubsection{Comparison with \citet{McConnachie2007}}

Although the derived total density profile is similar to \citet{McConnachie2007}, the old and intermediate
stellar density profiles are different. \citet{McConnachie2007} found a density profile flat in the center with  a
Sersic index of 0.3 for the old stellar population. We have found an exponential density profile for the old stars.
Concerning the intermediate-age stars, \citet{McConnachie2007} found an exponential profile which dominates the old
stars in the centre of the dwarf ($r < 500$ pc). Consequently, the intermediate-age stars represent more than 70\% of
the luminosity in the centre, therefore around 60-70\% of the mass if we take into account that their $M/L$ ratio is
larger than the one of the old stars. This is in contradiction with the fact that with the new SFH determination it is
now known that only 30\% of the stars have an intermediate age in the center of And II ($r < 500$ pc).

\subsubsection{Comparison with simulations}

\begin{figure}
  \begin{center}
    \centering
    \includegraphics[width=1\linewidth]{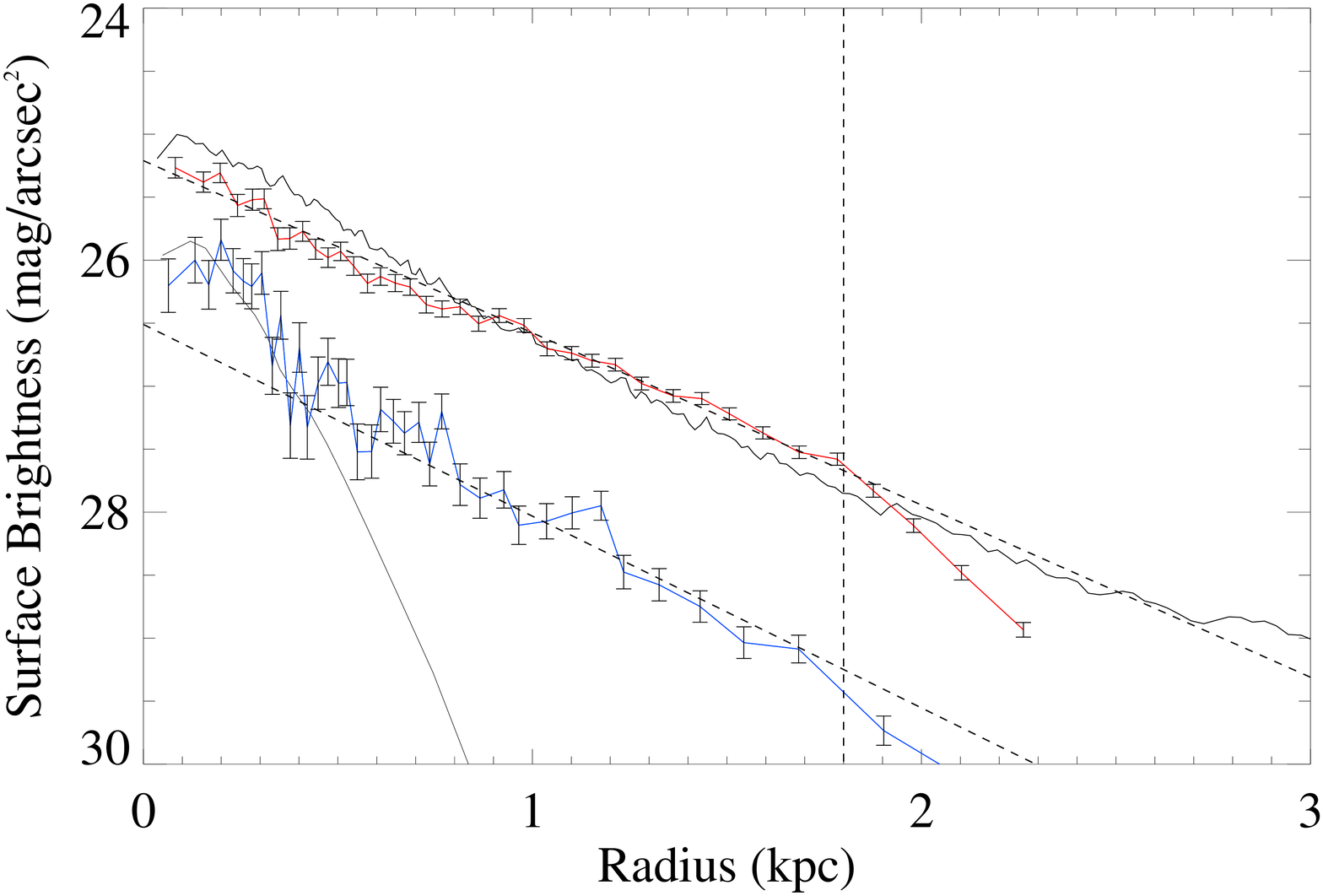} \\
    \includegraphics[width=1\linewidth]{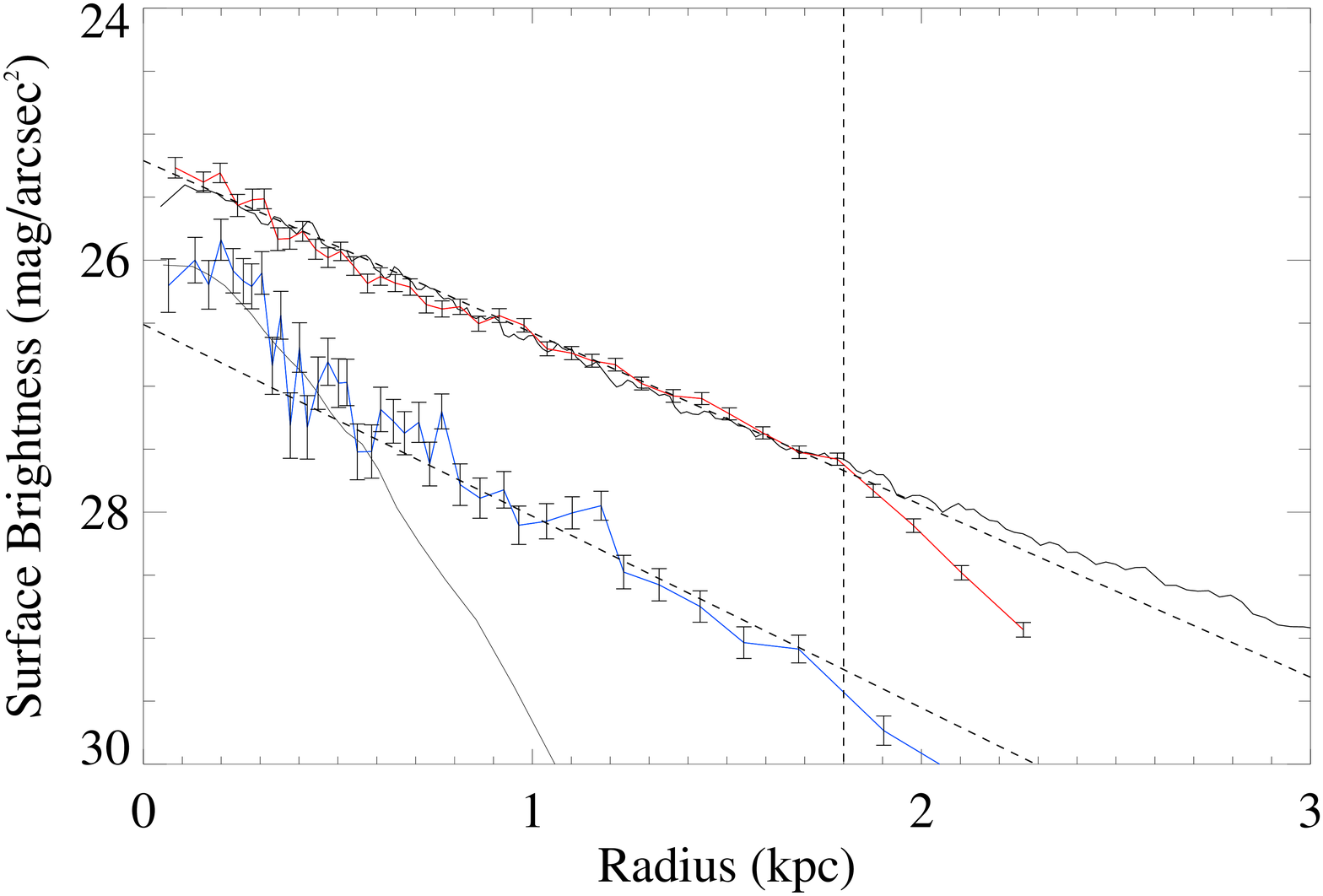}
    \caption{Comparison of the observed and simulated density profiles. Top panel:
before the ram pressure stripping. The density profile of the observed old stars (red line with error bars) is less
steep compared to the one from the simulation (black line) within 1.8 kpc (vertical lines). Within 0.5 kpc,
the density profile of the observed intermediate-age stars (blue line with error bars) is similar to the one in the simulation 
(black line). Beyond this radius, it seems to be contaminated by old stars. Bottom panel: one Gyr after the complete 
gas stripping. The slope of the profile for the old stars has changed and the fit is better. Intermediate-age stars are 
more extended and fit observations within 0.6 kpc.}
\label{fig:res_majmer_denspro}
  \end{center}
\end{figure}

We have arbitrarily rescaled the observed density profile for which the unit is star counts and not surface
brightness due to the fact that the mass-to-light ratios are not well constrained. For the primordial stars,
observations show a part of the curve ($0< r < 1.8$ kpc) which can be fitted by an exponential density profile, a
straight line, with a slope of 1.52 mag arcsec$^{-2}$ kpc$^{-1}$ (dashed lines in Figure \ref{fig:res_majmer_denspro}).
Beyond 1.8 kpc, the density drops very quickly. This decrease can be due to the cut in the stellar profile (a physical
explanation) or the lack of data (an instrumental explanation). For the simulation, the density profile of old stars
also follows a straight line (from 0 to 1.8 kpc) with a similar slope of 1.68 mag arcsec$^{-2}$ kpc$^{-1}$. The
difference with respect to the observations is about 10\%.

For the intermediate-age stars the errors are larger but it seems that a straight line can be a good
approximation of the density profile especially at radii between 0.5 and 1.8 kpc. In fact, this portion of the
intermediate-age stars can be well fitted by the same line as for the old stars with a smaller scale, $-1.36$ mag
arcsec$^{-2}$ and shows the same density drop after 1.8 kpc. This confirms the expectations from del Pino et al. (in
preparation): at radii larger than $\sim$ 0.5 kpc there is no more intermediate-age stars, this stellar population is
contaminated by the old stars. On the contrary, within $\sim$ 0.5 kpc the density profile is different from the one for
old stars with a bump in density which is well reproduced by the simulation (see Figure \ref{fig:res_majmer_denspro}).

\subsection{Stellar kinematics}

In the initial conditions, the two progenitors have angular momenta of the same magnitude ($2.49 \times 10^8$
\msun kpc km s$^{-1}$) but with different directions. The total angular momentum along the $x$-axis is equal to
$3.52 \times 10^8$ \msun kpc km s$^{-1}$. After the major merger, 5 Gyr ago, the angular momentum of the stars points 
towards the $x$-axis as expected from the conservation of this quantity along the symmetry axis: its direction is almost 
exactly along the $x$-axis and its magnitude is $3.42 \times 10^8$ in the same units, very close to the expected value 
(the difference being only 3\%).

\begin{figure}
  \begin{center}
    \centering
    \includegraphics[width=0.8\linewidth]{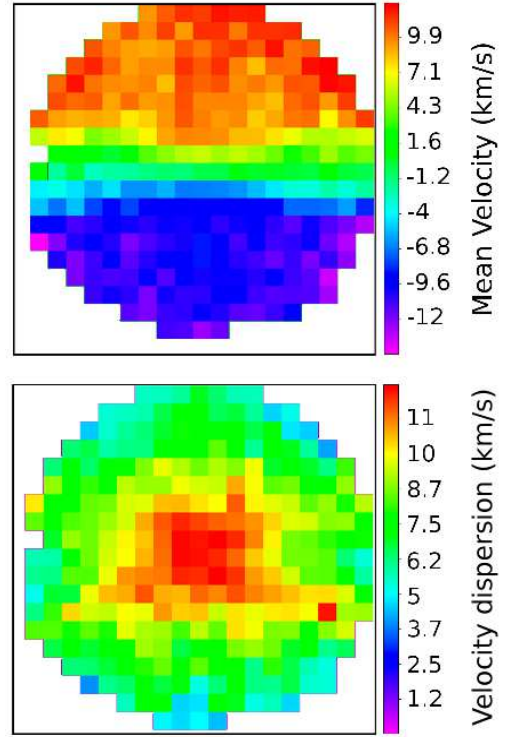}
    \caption{Stellar kinematics maps of the major merger remnant in a frame of 8 kpc $\times$ 8 kpc with a
resolution of 0.4 kpc in the $xy$ plane. The colour coding of the values is indicated on the right hand side of each
picture. The top panel shows the map of the mean velocity of the stars while the bottom panel presents the map of the
stellar velocity dispersion.}
\label{fig:res_majmer_kine}
  \end{center}
\end{figure}

Figure \ref{fig:res_majmer_kine} presents the mean velocity and velocity dispersion of
the stars in the remnant galaxy 5 Gyr ago within a projected radius of 4 kpc using bins of 0.4 kpc in the $xy$ plane.
This map shows clear rotation signal of stars around the $x$-axis as expected with this value of the angular
momentum.

\begin{figure*}
  \begin{center}
    \centering
    \includegraphics[width=0.45\linewidth]{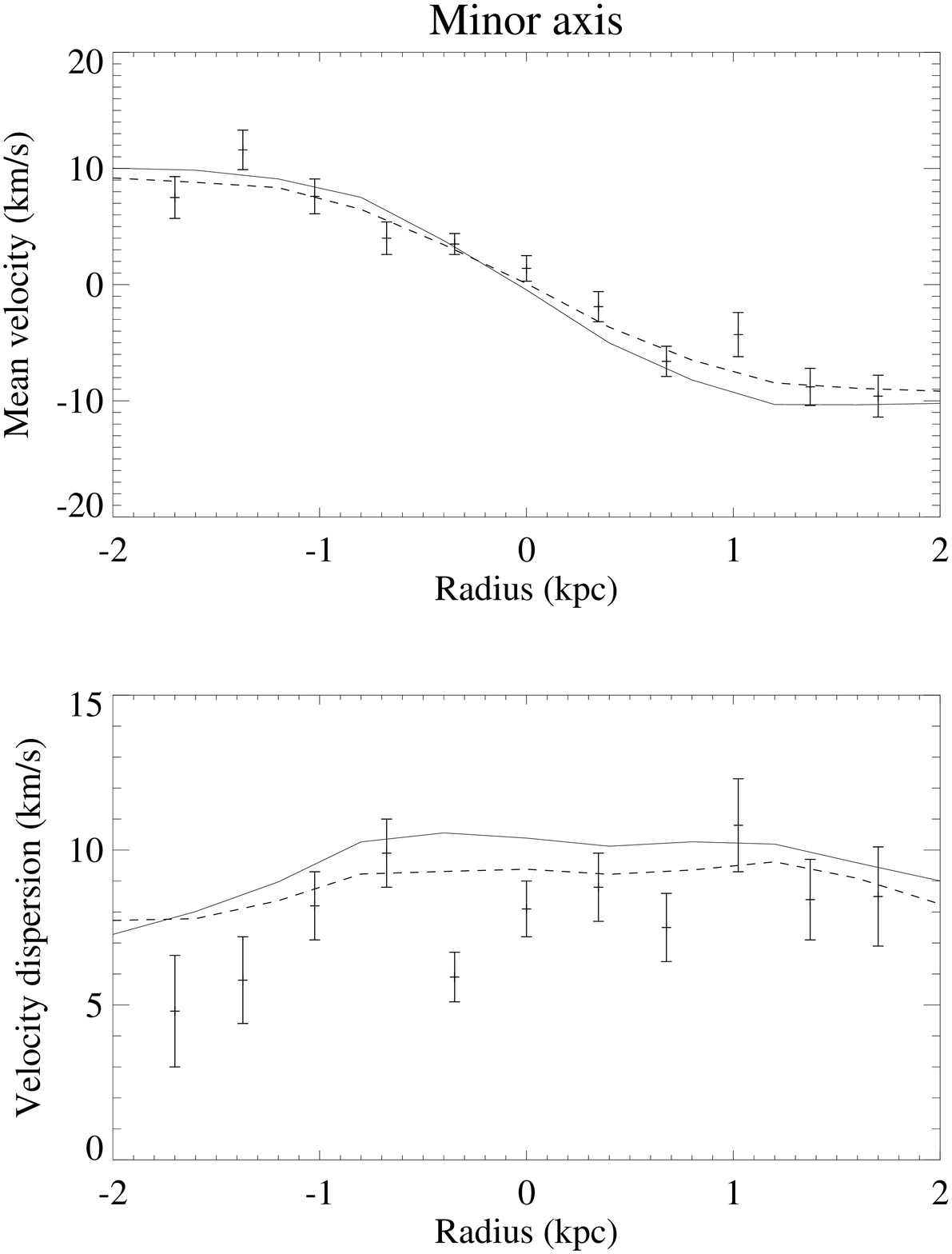}
    \includegraphics[width=0.45\linewidth]{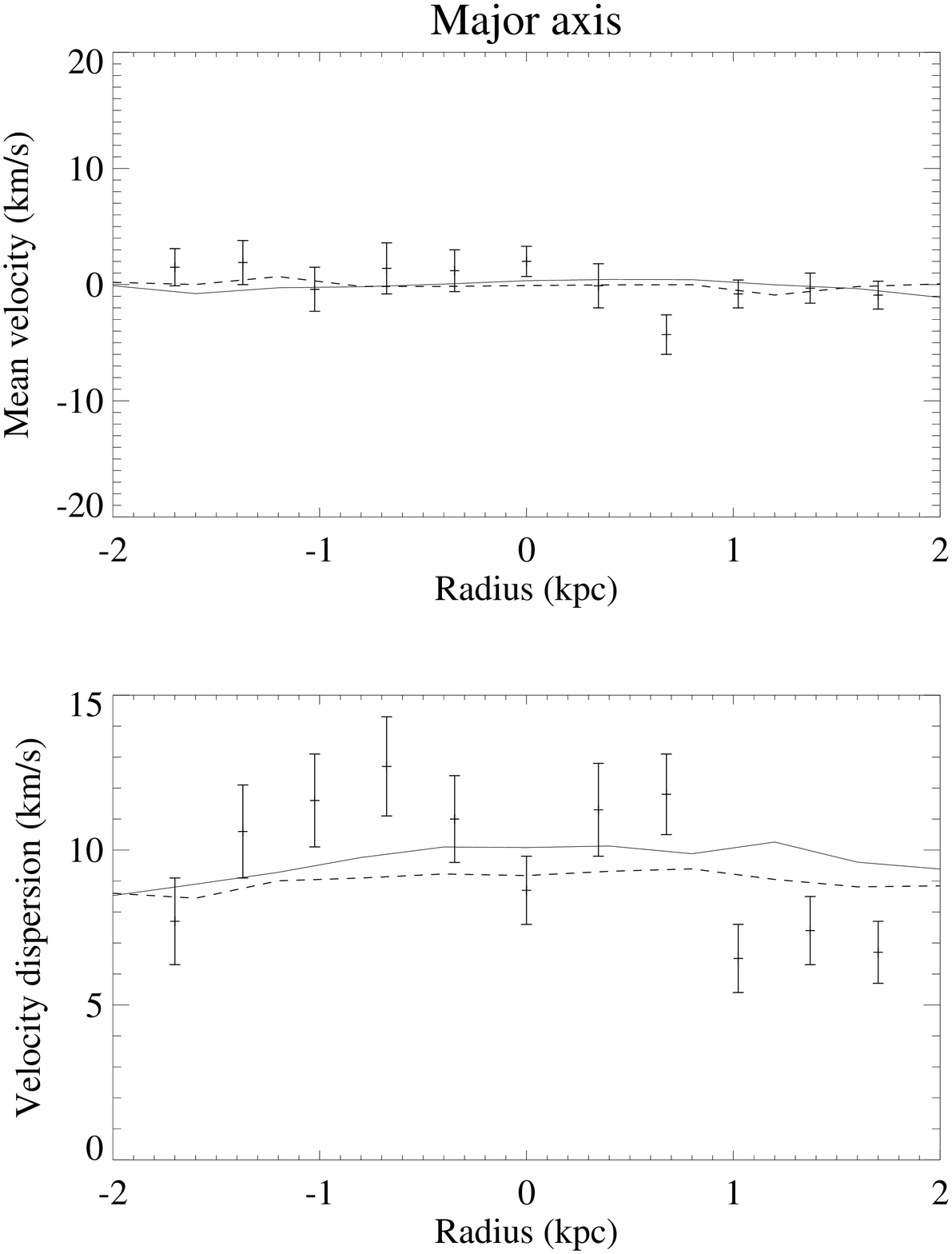}
    \caption{Comparison of 1D kinematics between the observations \citep[][points with error bars]{Ho2012} and
the simulations (solid and dashed lines). The measurements from the simulation were done in bins of size equal to 0.4
kpc, close to the irregular observational bins. The solid lines represent the measurements after the merger but before
the gas stripping and the dashed ones after the gas is stripped. The left-column plots show the measurements along the
minor axis of the galaxy image, the right-column ones those along the major axis. The plots in the upper row show
results for the mean velocity, those in the lower row for the velocity dispersion.}
\label{fig:res_majmer_kine_2}
  \end{center}
\end{figure*}

We have also calculated the mean velocity and velocity dispersion within 2 kpc
along the minor and major axis in order to mimic the observations presented in \cite{Ho2012}. Figure
\ref{fig:res_majmer_kine_2} shows that the simulated mean velocity along the minor and major axis reproduces within
1-1.5 $\sigma$ error bars the observed values except for one point. Concerning the velocity
dispersion, the simulation agrees with the observations within $2\sigma$ except for three data points among 22 for the
minor and major axis and agrees within $1 \sigma$ for 11 points out of 22.

\subsection{Properties of the gas component}

At the beginning of the simulation, the gas comprises $\sim 67$\% of the baryonic mass (baryonic mass = $60 \times
10^6$ \msun, gas mass = $42 \times 10^6$ \msun, stellar mass = $18 \times 10^6$ \msun). After the fusion, the gas
starts to be turned into stars. Until 5 Gyr ago, $1.8 \times 10^6$ \msun\ of the gas has been transformed into stars.
In addition some of the gas and stars were ejected due to the major merger. The visible stellar mass (above the
surface brightness of 28 mag arcsec$^{-2}$) is roughly $12 \times 10^6$ \msun$\,$ and the one of the gas is
$30 \times 10^6$ \msun\, corresponding to column density of $10^{19}$ to $10^{21}$ atoms cm$^{-2}$. The gas fraction
represents 71\% of the baryonic mass, i.e. it is approximately the same as before the merger.

The gas column densities and the gas total mass are large enough to be detected by radio telescopes as was the case for
other dwarf galaxies in the LG with HI masses below $10^7$ \msun$\,$ \citep{McConnachie2012}. And II seems to contain
no gas like the other MW dSph galaxies \citep{Spekkens2014}. Consequently, our scenario so far missed the physical
processes which would allow us to remove all the gas. In the next section, we investigate the interaction between M31
and the merger remnant as a possible mechanism to remove the gas and stop star formation.

\section{Interaction with M31}

And II is likely a satellite of M31. In the previous sections, we did not take into account the possible interaction
between And II and M31 and its consequences. As we assume the gas-rich major merger occurs in the field far from M31
(9-10 Gyr ago), the interaction with M31 has to occur later. We assume it occurs 5-7 Gyr ago to be consistent
with the star formation history which ends 5 Gyr ago. In this section, we use the snapshot which corresponds to the And
II state 5 Gyr ago when the star formation ends.

\begin{figure}
  \begin{center}
    \centering
    \includegraphics[width=0.89\linewidth]{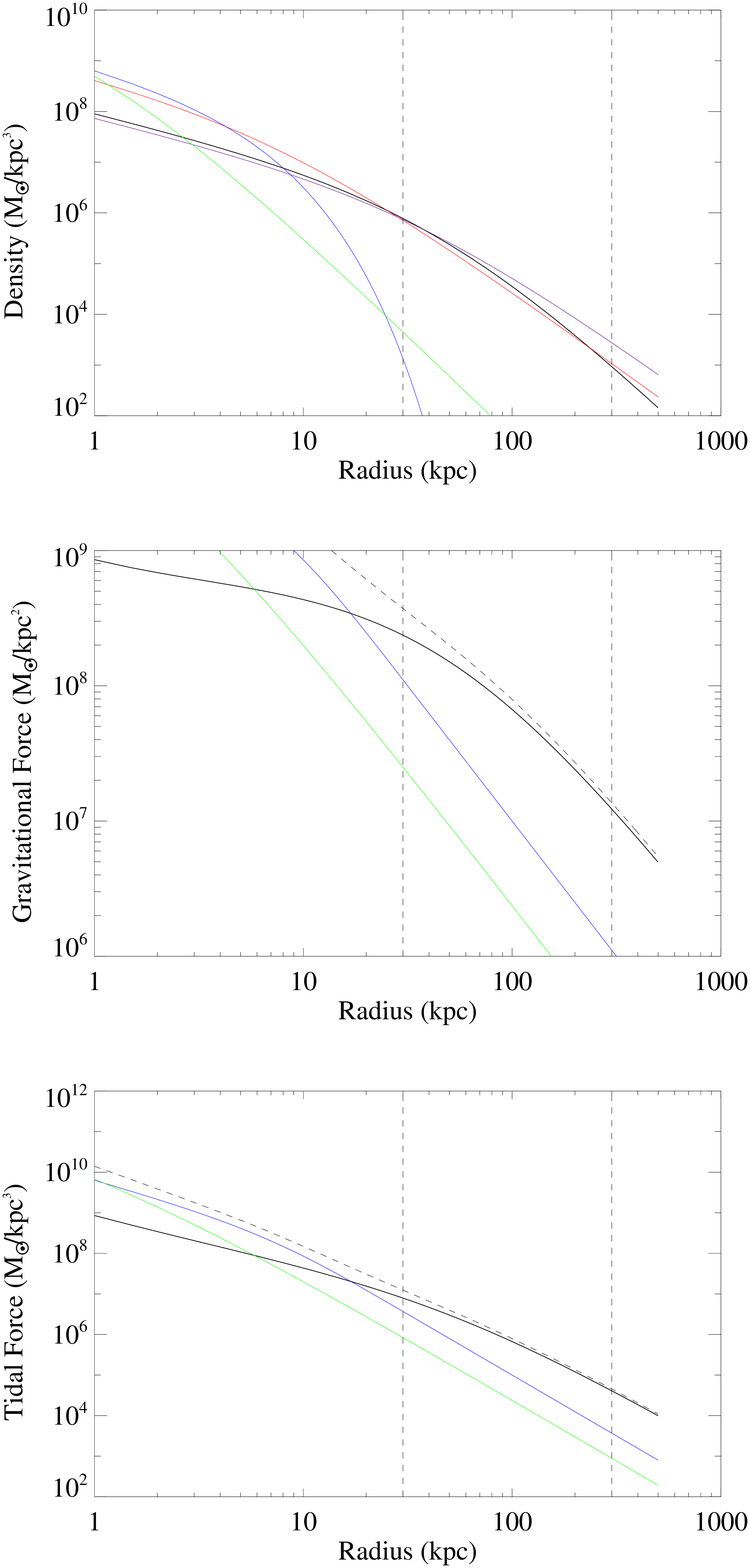}
    \caption{Top panel: the 3D density profile of the M31 model composed of the disk (blue), the bulge (green) and
the dark matter halo (black). Two additional lines correspond to other determinations of the dark matter density
profile: the purple line has parameters from \citet{Sadoun2014} (NFW profile with: $r_s$ = 7.63 kpc, $M_{200} = 0.88
\times 10^{12}$ \msun, $c = 25.5$, $R_{200} = 195$ kpc) and the red line from \citet{Sofue2015} (NFW profile with: $r_s
= 34.6$ kpc, $\rho_0 = 2.23 \times 10^{-3}$ \msun pc$^{-3}$, $M_{200} = 1.23 \times 10^{12}$ \msun). The vertical dashed
lines indicate the radial range between 30 and 300 kpc where And II is assumed to orbit. Middle panel: an approximation
for the gravitational force from the M31 components: $M(<r)/r^2$. Bottom panel: an approximation for the tidal force
generated by the same components: $M(<r)/r^3$. The dashed curve in the middle and bottom panel corresponds to the
total mass (baryons + dark matter).}
\label{fig:M31_mass_dens}
  \end{center}
\end{figure}

\subsection{The model of M31}

M31 is assumed to be composed of a bulge, a disk and a dark matter halo. The dark matter halo is a stable spherical
collisionless system for which velocity dispersion balances gravitational attraction. Its mass is set at $1.5 \times
10^{12}$ \msun$\,$ in agreement with the results from \cite{Watkins2010}. Its shape is assumed to be given by a
Hernquist profile with a scale radius of $r_{\rm a} = 50$ kpc. These parameters are chosen to be consistent
with other studies which have investigated the density profile of M31 between 30-300 kpc where And II is assumed
to orbit (see Figure \ref{fig:M31_mass_dens}).

The disk is modelled by an exponential disk which includes stars and gas. Its mass is taken to be $8 \times 10^{10}$
\msun\, and its scale radius is 3 kpc. The bulge is modelled by a Hernquist profile with a mass of
$2.5 \times 10^{10}$ \msun\, and a scale length of 1 kpc \citep{Widrow2003}.

If all the components are assumed to be spherical, even the disk, their cumulative masses contribute to
their gravitational forces as $\propto M(<r)/r^2$. The tidal forces they would exert on And II are $\propto
M(<r)/r^3$. Figure \ref{fig:M31_mass_dens} shows that between 30 and 300 kpc, the dark matter component provides a
dominant contribution to the cumulative mass, thus the gravitational force and the tidal force.
Consequently, in the simulation presented in the next section, where And II is always at a larger distance from M31
than 30 kpc, M31 is only modelled analytically by its dark matter halo.

\subsection{The orbit of the merger remnant around M31}

In order to compute the past orbit of And II around M31, we would need to know its present 3D distance to M31, its 3D
velocity with respect to M31 and the mass and density profile of M31 during the last few Gyr. What we actually know
is the 3D distance of And II and M31 \citep{Conn2011}, the radial velocities of M31 and And II in the MW rest frame and 
approximately the mass and the density profile of M31 (see Figure \ref{fig:M31_mass_dens}). We
therefore have to make some assumptions to derive the orbit of And II in the past. We assumed (1) that the M31
density profile has not changed during the last 6 Gyr, (2) that tangential velocities of M31 and And II
are smaller than 200 km s$^{-1}$ \citep{Carlesi2016}, (3) that the And II trajectory takes it
close to M31 $\sim$ 5 Gyr ago, to a distance between 30 and 40 kpc, to match the end of star formation 5 Gyr ago.
This range is chosen in order not to destroy the dwarf and to agree with our model of M31 (the pericentric distance
$r_{\rm p} > 30$ kpc) but at the same time to affect And II enough to remove its gas ($r_{\rm p} < 40$ kpc).

\begin{figure}
  \begin{center}
    \centering
    \includegraphics[width=1\linewidth]{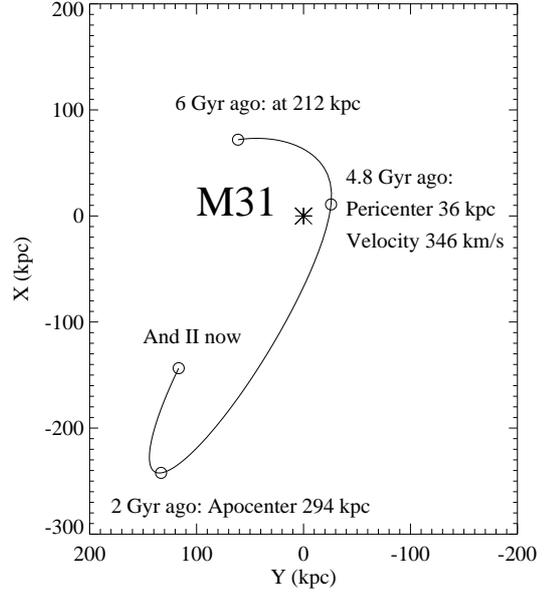}
    \caption{The trajectory of And II during the last 6 Gyr in the outskirts of M31. The $xy$ plane is defined by
the present position of And II, M31 and the MW. The plot is centred on M31. The $x$-axis corresponds to the direction
MW-M31 and the $y$-axis is the projected direction M31-And II. The trajectory of And II does not lie in this plane. The
most important moments of the orbit are labelled.}
\label{fig:M31_orbit}
  \end{center}
\end{figure}

By choosing randomly tangential velocities for And II and M31 within the limit of 200 km s$^{-1}$, we find that there
is only one family of trajectories that fit our assumptions. These trajectories turn out to have only one
pericentre close to M31. These pericentres can occur between 3.2 and 5.8 Gyr ago. We have selected pericentres to be
between 30 and 40 kpc and then the apocentres lie between 330 and 360 kpc. The properties of the trajectories are
mainly constrained by the constraint on the pericentre.

Figure \ref{fig:M31_orbit} shows one of the trajectories
we have selected. For this trajectory, 6 Gyr ago And II was at a distance of 212 kpc from M31 with a relative
velocity of 118 km s$^{-1}$. The pericentre passage happened 4.8 Gyr ago at a distance of 36 kpc with a velocity of 346
km s$^{-1}$ and the apocentre is at 294 kpc. In addition to the trajectory choice, we can choose the orientation of the
remnant galaxy. A purely retrograde interaction, when the direction of the dwarf's angular momentum is opposite to the
orbital angular momentum, reduces the internal stirring and mass loss \citep{Lokas2015}. This choice allows us to
maintain the internal kinematics of the dwarf remnant which fits the observations but limits the gas ejection. The
inverse, an exactly prograde interaction, has opposite effects.

\subsection{Tidal stripping}

We launch two simulations, one prograde and one retrograde, which represent two extreme cases, in order to
check the range of influence of tidal stripping on And II. In each simulation, star formation is turned off to make
them computationally faster and study only one effect at a time. We use the remnant galaxy from the previous
merger simulation corresponding to And II 5 Gyr ago at the expected position of And II 6 Gyr ago (see the previous
section) and we follow the simulation for 6 Gyr, i.e. until the present time. The snapshot choice, 5 Gyr ago instead of
6 Gyr ago, was done in order to have the right amount of new stars for And II. There is no star formation during the
tidal stripping simulation. Moreover, 5 or 6 Gyr ago, And II had similar properties which do not change the
main result of the simulation.

After 6 Gyr, there remains  93\% (retrograde case) and 88\% (prograde case) of the initial
gas in the dwarf. The gas stripping due to the close encounter with M31 turns out to be insufficient. It is mainly due
to the high density of the dark matter halo and the baryonic matter in the dwarf galaxy centre. The difference
due to the orbital orientation takes the form of a nearly twice longer tail and a larger fraction of stripped mass
for the prograde with respect to the retrograde trajectory (12\% instead of 7\%).

We conclude that the tidal stripping cannot be the only mechanism of gas stripping. Moreover, if the gas could be fully
stripped by tidal forces, it would imply that the tidal force was strong enough to also strip the stars and the dwarf
galaxy would be completely or partially destroyed. The only way to remove the gas without destroying the whole
dwarf galaxy is to employ a process which affects just the gas and not the stars: the ram pressure stripping. Finally,
if the tidal force does not affect the gas in the central parts it does not affect the stars neither. This means that
the stellar properties in terms of morphology and kinematics are conserved.

\section{Ram pressure stripping by M31 halo}

In addition to the disk, bulge and dark matter halo, galaxies like the MW or M31 seem to have a hot gas halo
\citep{Miller2015}. These haloes are believed to be responsible for a well known observational fact in the local
Universe: dwarf galaxies in the vicinity of massive galaxies or clusters have no gas, while dwarf galaxies in the field
have large gas fractions \citep{Dressler1980, Spekkens2014}. This fact is explained by the mechanism of
ram pressure stripping which reduces the gas content in dwarf galaxies as a result of interaction with the hot gas of 
their host \citep{Gunn1972}.

The process depends mainly on two parameters: the density of the hot gas ($\rho$) and the velocity of the dwarf galaxy
moving within this medium ($v$). The ram pressure is proportional to $\rho v^2$.
In our study, in order to have efficient ram pressure stripping, the dwarf remnant of the major merger must
be on an eccentric orbit and pass close to the centre of M31 where the gas density and the dwarf velocity are the
highest. These assumptions are consistent with the type of orbit described in the previous section.

\subsection{The hot gas halo of M31}

The hot gas halo of M31 is assumed to be spherical without streaming velocity nor velocity dispersion, i.e. we set $v
= 0$ for all gas particles in the initial conditions. The pressure supports gravity only by the internal temperature
which is very high $\sim$ 10$^6$ K. There are no strong observational constrains concerning the mass and density
distribution of the hot gas halo of M31. However, previous works on hot gas haloes seem to indicate that these
structures have mass roughly 100 times smaller than the dark matter mass of the host halo and have a similar spatial
distribution \citep{Miller2015}. Consequently, we approximate the hot gas halo of M31 as a Hernquist profile with a
total mass of $1.5 \times 10^{10}$ \msun$\,$ and a scale length of 50 kpc, as for the dark matter halo. At 36 kpc, the
assumed pericentre of the orbit, the density reaches $10^4$ \msun kpc$^{-3}$ corresponding to $4 \times 10^{-4}$
atoms cm$^{-3}$.

\subsection{Simulation in a periodic box}

In order to simulate the complete interaction between M31 and the remnant of the dwarf
galaxy major merger with a constant gas resolution, we would need about 80 million particles for the hot gas halo, which
is too expensive computationally. In order to reduce the simulation time and at the same time reliably model the main
physical process, we use a periodic box. We fill the box with hot gas of constant density of $4 \times 10^{-4}$
atoms cm$^{-3}$ and temperature of $10^6$ K. The box has a size of 30 kpc in order to be significantly larger than the
dwarf galaxy itself ($> 10$ times the half-light diameter) and at the same time not to contain too many particles
(close to 1 million). The remnant dwarf galaxy is placed at the centre of the box without changing the orientation of
its angular momentum. We set a rather high bulk velocity of the dwarf $v_x = 400$ km s$^{-1}$.
The chosen density and velocity are consistent with the conditions at the pericentre discussed in the previous section.
The velocity vector of the galaxy is parallel to the direction of the angular momentum which helps to
remove the gas.

We have decided to not include star formation in this ram-pressure stripping simulation. However, we have ran a
test simulation with star formation and found that, as expected, ram pressure compresses the gas and thus increases the
star formation. This star formation process does not form a stellar population that would be considered young at
present because the ram-pressure stripping due to M31 is assumed to occur before 5 Gyr ago and to finish nearly 5 Gyr
ago. Due to ram pressure the star formation can become stronger by a factor of 2-3. At the same time, ram pressure
reduces the amount of gas and thus decreases the global star formation. In our ram-pressure model, we try to reproduce
only the period when the ram-pressure stripping is very efficient thus when the dwarf is very close to M31 (30-60 kpc)
so that the dwarf velocity and the hot gas halo have their highest values. This period is short, less than 1 Gyr.
Consequently, the star formation due to the ram pressure would just increase the star formation during a small period
before the complete stripping of the gas. The amount of stars due to the ram pressure would represent less than 15\% of
the total mass of stars of the second stellar population. This is therefore an effect of second order on the star
formation. Moreover, we do not take into account ram-pressure stripping when the simulated dwarf is further from M31
(60-150 kpc). In this region, ram-pressure stripping is less efficient but lasts longer. It is probably not
sufficiently strong to increase the central density and star formation but it can remove the gas in the outer parts
and reduce the remaining `gas fuel'. We think that only a complete simulation, including all the processes
simultaneously, can properly model the star formation process when And II is on an orbit around M31.

\subsection{Results of the ram pressure stripping}

\begin{figure}
  \begin{center}
    \centering
    \includegraphics[width=0.8\linewidth]{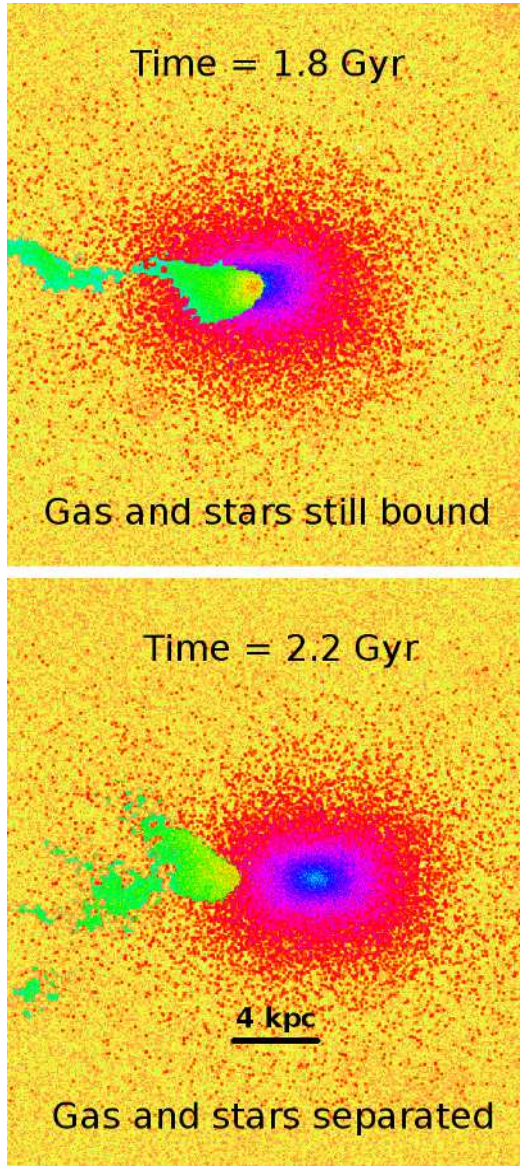}
    \caption{Ram pressure stripping of the gas in the merger remnant by the hot gas halo of M31.
In the background there is the hot gas halo of M31 with a quasi-constant density shown in yellow colours. The
red to blue colours correspond to the stars of the dwarf which retain their shape in time and are not affected by the
hot gas. Green colours indicate the gas of And II which is elongated along the $x$-axis because the dwarf's
velocity, thus the ram pressure stripping, is along this axis. The stars and gas separate between 1.8 (upper panel) and
2.2 (lower panel) Gyr of simulation in the periodic box. At 2.2 Gyr the gas is fully stripped and forms a clump 4
kpc behind the stars.}
\label{fig:res_RPS}
  \end{center}
\end{figure}

After 0.5 Gyr of evolution in the periodic box with a constant velocity and a high hot gas halo density, the gas
remaining around the dwarf galaxy centre represents less than 60\% of the initial gas. After about 2 Gyr, the gas and
stars start to separate. After 2.2 Gyr, they are completely disconnected and the dwarf becomes gas free (see Figure
\ref{fig:res_RPS}). This proves that the gas can be fully stripped from And II via ram pressure stripping. However,
this simulation requires too much time (more than 2 Gyr) with extreme conditions ($v_x = 400$ km s$^{-1}$ and
$\rho = 4 \times 10^{-4}$ cm$^{-3}$). For a realistic orbit, And II would remain close to the pericentre (between 70
and 30 kpc) during only $\sim$ 0.5 Gyr. How can we reduce the gas stripping time? It does not seem reasonable to
increase the velocity even more because And II would then be unbound from M31. On the other hand, the density of the
hot gas could be larger because the mass of the hot gas halo is not strongly constrained. If the gas density is
increased by a factor of two or three, the stripping time decreases to values as low as 0.7 Gyr and 0.3 Gyr,
respectively, for the same simulation.

Our simulation is obviously simplified, it does not include star formation which decreases the
central density and thus facilitates the ram pressure stripping, nor does it take into account the tidal force
which would elongate the gas distribution making it easier to strip. However, the central density of the gas
with or without star formation differs only by a factor of about 2 and the star formation occurs only at the
centre of the dwarf galaxy. Concerning the tidal force, it is mostly efficient in the outer parts. Another way to
obtain faster gas stripping would be to modify the dark matter density profile. Indeed, the force which resists the
ram pressure stripping is the gravitational force mainly due to dark matter. For comparison, the mass that
generates the gravitational force within 0.3 kpc is $8.6 \times 10^6$ \msun\ for the dark matter, $0.78 \times 10^6$
\msun\ for the gas and $0.88 \times 10^6$ \msun\ for the stars. With a lower central density, the gas could be more easily
stripped. Among the simulations we ran, those with smaller dark matter content indeed could strip the gas within 1 Gyr
but the properties of the remnant dwarf did not match the observations so well. This is consistent with the recent study by 
\citet{Emerick2016} which shows that changing a cusp into a core in the dark matter profile, i.e. decreasing the central dark 
matter density, helps to remove the gas.

\subsection{Stellar properties after the ram-pressure stripping}

In addition to gas stripping, ram pressure has a small effect on the stars. The ellipticity slightly
increased from 0.22 to 0.29 at the new half-light radius of 1.0 kpc in the $xy$ plane, which differs by $1.2\sigma$
from the observations. This result can be due to the assumed alignment between the dwarf velocity and angular momentum.
The gas is thus ejected along the $x$-axis, the major axis of the dwarf, which can increase the extent of the dwarf in
this direction and the ellipticity. Figure \ref{fig:res_majmer_kine_2} illustrates the fact that the observed velocity
dispersion has decreased. This is due to the decrease of the mass of the dwarf and more specifically in the center
after the stripping of the gas. In parallel, the stellar central density has also slightly decreased. This leads to a
better fit to the observations for the density profiles of the old and intermediate-age stars (see Figure
\ref{fig:res_majmer_denspro}).

\section{Discussion}

\subsection{Complex histories of dwarf galaxies}

With the new studies concerning the SFH \citep{Weisz2014} and kinematics of dwarf galaxies \citep{Ho2012, dPM2016},
dSphs and dIrrs seem more complicated than they have ever been. DSphs were thought to be the descendants of
primordial disky galaxies accreted by a bigger host galaxy and transformed morphologically and kinematically
into pressure-supported systems over a time-scale of a few Gyr. This scenario may still be true for galaxies with only
one old stellar population as is the case for some ultra-faint dwarf galaxies. However, it does not hold for dSphs like
And II, Fornax, Leo I and Carina which seem to have complicated and extended SFHs \citep{Weisz2014}. For example, Leo I
was still forming stars 1-2 Gyr ago.

In this context, And II appears to be an example of a dwarf with a complicated history. First, it probably is a result
of a major merger between two gas-rich dwarf galaxies. Second, it could have been quickly stripped of its
gas 5 Gyr ago after one close passage around M31. In the LG, massive dwarf ellipticals or dSphs like NGC205, NGC147,
M32 or Fornax, show either rotation or complicated kinematical structure \citep{Geha2006, dPM2016} and some
even retain some gas, e.g. NGC205 which is much more massive than And II \citep{Young1997}. Each dwarf galaxy seems to
be a specific and complicated puzzle to solve.

\subsection{Observational test of the model}

The first and most important test of our model would be provided by the velocity of And II in the M31 rest
frame. When the tangential velocities of M31 and And II are measured with high enough accuracy, they should indicate a
close passage around M31 (at a distance of 30-50 kpc) in the past (4-6 Gyr ago). If it turns out instead that And II's
does not follow a highly eccentric orbit, it will mean that the gas stripping is not due to the interaction with M31.
Nowadays, even the tangential velocity of M31 is very uncertain. Two recent measurements give very different,
inconsistent results: a value of 17 km s$^{-1}$ from \cite{Sohn2012} and 164 km s$^{-1}$ by \cite{Salomon2016}. For And
II the tangential velocity is completely unknown.

Another test could be the presence of High-Velocity Clouds (HVC) as the result of the quick stripping of gas
from And II 5 Gyr ago. Our simulations (see Figure \ref{fig:res_RPS}) and those of \citet{Mayer2007} show that when a
galaxy is stripped of gas by ram pressure some of the stripped gas can form gas clouds. A new HI survey
\citep{Kerp2016} of the outskirts of M31 has recently provided some hints that an elongated HVC observed there could be
formed from the gas stripped from Andromeda XIX. The HVCs between And II's position and the centre of M31 could also be
the traces of the stripping. However, even if the hypothesis is true, it would be difficult to prove because in our
scenario the gas has been decoupled from the dwarf for the last $\sim 5$ Gyr and the two components must have followed
different trajectories in the hot gas halo. A full simulation of the interaction of And II with the M31 hot gas halo
including ram pressure stripping, the formation HVC and tracing its trajectory could provide more insight.

\subsection{Removal of all the gas by star formation}

How is it possible to strip And II's gas in one passage close to M31? One mechanism, that of ram pressure stripping was
discussed in the previous section. Another possibility could be the star formation if we assume that during the major
merger all the gas is turned into stars. This would mean that the gas mass during the fusion is the same as the mass
of intermediate-age stars: roughly $1.8 \times 10^6$ \msun. This implies an initial gas fraction close to 20\% or
a bit larger ($\sim$ 30\%) if we take into account the gas ejected during the major merger, which seems to be too
low given the observational constraints. Moreover, from the theoretical point of view it is difficult to consume all
the gas in a dwarf galaxy only by the star formation process. This would require to have a star formation activity even
when the gas density is low, i.e. below the commonly accepted density
threshold.

The other possibility is to remove the gas by supernovae. An interesting example in this context is a galaxy like Leo P 
which is full of gas with star formation going on only at the centre where the density is high enough \citep{McQuinn2015}. 
However, this level of star formation ($10^{-5}$ \msun yr$^{-1}$) is not high enough to blow away the gas and if the gas 
density is decreased further due to the star formation, the dwarf will not be able to form new stars. It would remain a dwarf 
galaxy with a few stars and a large reservoir of gas, not a dSph. However, since Leo P is more than 10 times less massive than 
And II, the comparison may be not completely relevant. This conclusion is supported by \citet{Emerick2016} who found that 
gas stripping by supernova feedback is not very efficient.

\subsection{Limitations of the model}

The first limitation of our model of And II is its degeneracy. In order to reduce the number of parameters we have
assumed that the two dwarf galaxies are exactly the same. It is a useful simplification but it indicates that the
problem is degenerate. Moreover, even with the same progenitors there are still degeneracies present between the
parameters. For example, the gas fraction and feedback both have an effect on the SFH. It is
possible to choose a different pair of values for the feedback and gas fraction parameters and get a very similar dwarf
remnant. However, even if the parameters are allowed to change in our model, they cannot change too much.
If we want a gas-rich major merger to reproduce the observables the second progenitor cannot be too small or contain a
low gas fraction.

Second, another limitation concerns the inclination of And II. We assume implicitly in this study that And II is seen along
the line of sight which is perpendicular to its angular momentum vector. Contrary to disky galaxies, where the
projected shape allows us to infer inclination, there is no simple way to find the inclination of And II. It is quite
possible that And II is seen at a slightly different angle and its rotation is therefore even higher.

Third, our model depends on the exact numerical implementation of gas physics. This physics is much more complicated
to simulate than the stellar physics. For example, the effects of feedback and ram pressure stripping are known
to depend on the versions of code used \citep{OShea2005, Agertz2007}. With a more complete description of the baryonic
processes, the gas could be stripped faster and the star formation could eject more gas from the dwarf.

Fourth, the simulation of the ram-pressure stripping is idealized. The hot gas density and the dwarf velocity
are constant which is surely not true if And II was orbiting M31 on an eccentric orbit. Moreover, we do not include star
formation in this period. These drawbacks are directly linked to the fact that our model is constructed from two
simulations and not one complete simulation due to numerical limitations.

\section{Conclusions}

In this study, we have proposed a detailed scenario to explain the main properties of And II. In our model, this dSph,
a satellite of M31, is the result of a major merger between two gas-rich disky dwarf galaxies in the field at
high redshift ($z > $ 2-3). This major merger would have formed a new stellar population due to the gas compression
during the fusion phase. It would also have transformed the morphology of disky progenitors into a spheroidal remnant.
Due to the specific inclination of the angular momenta of the progenitors, the remnant galaxy would have kept some
rotation, but around the major axis and not the minor one as in disky galaxies.

In addition to this major merger, our model requires And II to fall into the potential well of M31
about 6 Gyr ago and reach the pericentre of its orbit $\sim$ 5 Gyr ago. As a result of this close interaction, And
II's gas would have been quickly removed mainly by ram pressure stripping but also by tidal stripping. This quick
stripping can explain why And II has stopped its star formation $\sim$ 5 Gyr ago.

Our simulations succeeded in reproducing the different properties of And II: ellipticity, morphology, kinematics, SFH,
and the properties of the stellar populations. Only the quick gas stripping is not properly reproduced. It could be due
to several reasons: (1) a problem in the initial conditions where we assume a too concentrated dark matter halo;
(2) an over-simplified recipe for star formation which does not properly take into account the effect
of supernovae which could eject more gas from the dwarf galaxy; (3) a simplified simulation of gas stripping which
does not take into account the tidal force from M31 and star formation or (4) too small density of hot gas in M31. These
effects alone or in combination may explain the lack of gas in And II at present.

In general, this study of the particular case of And II indicates, as \citet{Benitez2016} have done for a
cosmological simulation, that a major merger of gas-rich dwarf galaxies can be an important physical process to
transform dIrrs into dSphs and to explain their complicated SFHs.

\section*{Acknowledgements}

This research was supported in part by the Polish National Science Centre under grant
2013/10/A/ST9/00023 and by the project RVO:6798581.
We are grateful to L. Widrow for providing procedures to generate $N$-body realizations
of galaxies for initial conditions.

\bsp

\label{lastpage}


\begin{thebibliography}{99}

\bibitem[Amorisco et al.(2014)]{Amorisco2014} Amorisco N.~C., Evans N.~W., van de Ven G., 2014, \nat, 507, 335
\bibitem[Agertz et al.(2007)]{Agertz2007} Agertz O. et al., 2007, \mnras, 380, 963
\bibitem[Bechtol et al.(2015)]{Bechtol2015} Bechtol K. et al., 2015, ApJ, 807, 50
\bibitem[Belokurov et al.(2006)]{Belokurov2006} Belokurov V. et al., 2006, ApJ, 647, L111
\bibitem[Belokurov et al.(2007)]{Belokurov2007} Belokurov V. et al., 2007, ApJ, 654, 897
\bibitem[Ben{\'{\i}}tez-Llambay et al.(2016)]{Benitez2016} Ben{\'{\i}}tez-Llambay A., Navarro J.~F.,
	Abadi M.~G., Gottl\"{o}ber S., Yepes G., Hoffman Y., Steinmetz M., 2016, \mnras, 456, 1185
\bibitem[Carlesi et al.(2016)]{Carlesi2016} Carlesi E., Hoffman Y., Sorce J.~G., Gottl\"{o}ber S., Yepes G.,
	Courtois H., Tully R. B., 2016, MNRAS, 460, L5
\bibitem[Conn et al.(2011)]{Conn2011} Conn A.~R. et al., 2011, ApJ, 740, 69
\bibitem[Cox et al.(2006)]{Cox2006} Cox T.~J., Jonsson P., Primack J.~R., Somerville R.~S., 2006, MNRAS, 373, 1013
\bibitem[Deason et al.(2014)]{Deason2014} Deason A., Wetzel A., Garrison-Kimmel S., 2014, \apj, 794, 115
\bibitem[Dekel \& Silk(1986)]{Dekel1986} Dekel A., Silk J., 1986, ApJ, 303, 39
\bibitem[del Pino et al.(2015)]{dPM2015} del Pino A., Aparicio A., Hidalgo S.~L., 2015, \mnras, 454, 3996
\bibitem[del Pino et al.(2016)]{dPM2016} del Pino A., Aparicio A., Hidalgo S. L., {\L}okas E. L., 2016, \mnras,
	submitted, arXiv:1605.09414
\bibitem[Dressler(1980)]{Dressler1980} Dressler A., 1980, \apj, 236, 351
\bibitem[Ebrov{\'a} \& {\L}okas(2015)]{Ebrova2015} Ebrov{\'a} I., {\L}okas E.~L., 2015, \apj, 813, 10
\bibitem[Emerick et al.(2016)]{Emerick2016} Emerick A., Mac Low M., Grcevich J., Gatto A., 2016, ApJ, 826, 148
\bibitem[Geha et al.(2006)]{Geha2006} Geha M., Guhathakurta P., Rich R.~M., Cooper M.~C., 2006, \aj, 131, 332
\bibitem[Governato et al.(2010)]{Gover2010} Governato F. et al., 2010, Nature, 463, 203
\bibitem[Governato et al.(2012)]{Gover2012} Governato F. et al., 2012, MNRAS, 422, 1231
\bibitem[Grcevich \& Putman(2009)]{Grcevich2009} Grcevich J., Putman M. E., 2009, \apj, 696, 385
\bibitem[Gunn \& Gott(1972)]{Gunn1972} Gunn J.~E., Gott J.~R., 1972, \apj, 176, 1
\bibitem[Hammer et al.(2010)]{Hammer2010} Hammer F., Yang Y.~B., Wang J.~L., Puech M., Flores H., Fouquet S.,
	2010, ApJ, 725, 542
\bibitem[Hernquist(1993)]{Hernquist1993} Hernquist L., 1993, \apjs, 86, 389
\bibitem[Ho et al.(2012)]{Ho2012} Ho N. et al., 2012, \apj, 758, 124
\bibitem[Hunter et al.(2012)]{Hunter2012} Hunter D.~A. et al., 2012, \aj, 144, 134
\bibitem[Ibata et al.(2007)]{Ibata2007} Ibata R., Martin N.~F., Irwin M., Chapman S.,
	Ferguson A.~M.~N., Lewis G. F., McConnachie A. W., 2007, ApJ, 671, 1591
\bibitem[Kalirai et al.(2010)]{Kalirai2010} Kalirai J.~S. et al., 2010, \apj, 711, 671
\bibitem[{Kazantzidis et al.}(2011a)]{Kazan2011a} Kazantzidis S., {\L}okas E. L., Callegari S., Mayer L.,
	Moustakas L. A., 2011a, ApJ, 726, 98
\bibitem[{Kazantzidis et al.}(2011b)]{Kazan2011b} Kazantzidis S., {\L}okas E. L., Mayer L., Knebe A., Klimentowski J.,
	2011b, ApJ, 740, L24
\bibitem[Kerp et al.(2016)]{Kerp2016} Kerp J., Kalberla P.~M.~W., Ben Bekhti N., Fl\"{o}er L., Lenz D., Winkel B.,
	2016, A\&A, 589, A120
\bibitem[Khochfar \& Burkert(2006)]{Khochfar2006} Khochfar S., Burkert A., 2006, \aap, 445, 403
\bibitem[Kirby et al.(2014)]{Kirby2014} Kirby E.~N., Bullock J.~S., Boylan-Kolchin M., Kaplinghat M., Cohen J.~G.,
	2014, \mnras, 439, 1015
\bibitem[{Klimentowski et al.}(2009)]{Kliment2009} Klimentowski J., {\L}okas E. L., Kazantzidis S.,
	Mayer L., Mamon G. A., 2009, MNRAS, 397, 2015
\bibitem[Klimentowski et al.(2010)]{Klimentowski2010} Klimentowski J., {\L}okas E.~L., Knebe A.,
	Gottl\"{o}ber S., Martinez-Vaquero L. A., Yepes G., Hoffman Y., 2010, \mnras, 402, 1899
\bibitem[Koposov et al.(2015)]{Koposov2015} Koposov S.~E., Belokurov V., Torrealba G., Evans N. W., 2015,
	ApJ, 805, 130
\bibitem[Laevens et al.(2015)]{Laevens2015} Laevens B.~P.~M., et al. 2015, ApJ, 813, 44
\bibitem[{\L}okas et al.(2011)]{Lokas2011} {\L}okas E.~L., Kazantzidis S., Mayer L., 2011, ApJ, 739, 46
\bibitem[{\L}okas et al.(2012)]{Lokas2012} {\L}okas E.~L., Majewski S.~R., Kazantzidis S., Mayer L., Carlin J. L.,
	Nidever D. L., Moustakas L. A., 2012, ApJ, 751, 61
\bibitem[{\L}okas et al.(2014)]{Lokas2014} {\L}okas E.~L., Ebrov{\'a} I., del Pino A., Semczuk M., 2014,
	\mnras, 445, L6
\bibitem[{\L}okas et al.(2015)]{Lokas2015} {\L}okas E.~L., Semczuk M., Gajda G., D'Onghia E., 2015, ApJ, 810, 100
\bibitem[Martin et al.(2009)]{Martin2009} Martin N.~F. et al., 2009, ApJ, 705, 758
\bibitem[Martin et al.(2013)]{Martin2013} Martin N.~F. et al., 2013, ApJ, 772, 15
\bibitem[{Mayer et al.}(2001)]{Mayer2001} Mayer L., Governato F., Colpi M., Moore B., Quinn T., Wadsley J.,
	Stadel J., Lake G., 2001, ApJ, 559, 754
\bibitem[Mayer et al.(2007)]{Mayer2007} Mayer L., Kazantzidis S., Mastropietro C., Wadsley J., 2007,
	Nature, 445, 738
\bibitem[McConnachie \& Irwin(2006)]{McConnachie2006} McConnachie A.~W., Irwin M.~J., 2006, \mnras, 365, 1263
\bibitem[McConnachie et al.(2007)]{McConnachie2007} McConnachie A.~W., Arimoto N., Irwin M., 2007, \mnras, 379, 379
\bibitem[McConnachie(2012)]{McConnachie2012} McConnachie A.~W., 2012, AJ, 144, 4
\bibitem[McQuinn et al.(2015)]{McQuinn2015} McQuinn K.~B.~W. et al., 2015, \apj, 812, 158
\bibitem[Miller \& Bregman(2015)]{Miller2015} Miller M.~J., Bregman J.~N., 2015, \apj, 800, 14
\bibitem[Navarro et al.(1996)]{NFW1996} Navarro J.~F., Frenk C.~S., White S.~D.~M., 1996, \apj, 462, 563
\bibitem[Nichols et al.(2015)]{Nichols2015} Nichols M., Revaz Y., Jablonka P., 2015, A\&A, 582, A23
\bibitem[O'Shea et al.(2005)]{OShea2005} O'Shea B.~W., Nagamine K., Springel V., Hernquist L.,
	Norman M.~L., 2005, \apjs, 160, 1
\bibitem[Rodrigues et al.(2012)]{Rodrigues2012} Rodrigues M., Puech M., Hammer F., Rothberg B., Flores H.,
	2012, MNRAS, 421, 2888
\bibitem[Sadoun et al.(2014)]{Sadoun2014} Sadoun R., Mohayaee R., Colin J., 2014, \mnras, 442, 160
\bibitem[Sakamoto \& Hasegawa(2006)]{Sakamoto2006} Sakamoto T., Hasegawa T., 2006, ApJ, 653, L29
\bibitem[Salomon et al.(2016)]{Salomon2016} Salomon J.-B., Ibata R.~A., Famaey B., Martin N.~F.,
	Lewis G.~F., 2016, \mnras, 456, 4432
\bibitem[Skillman et al.(2016)]{Skillman2016} Skillman E.~D. et al., 2016, ApJ, submitted, arXiv:1606.01207
\bibitem[Sofue(2015)]{Sofue2015} Sofue Y., 2015, \pasj, 198
\bibitem[Sohn et al.(2012)]{Sohn2012} Sohn S.~T., Anderson J., van der Marel R.~P., 2012, \apj, 753, 7
\bibitem[Spekkens et al.(2014)]{Spekkens2014} Spekkens K., Urbancic N., Mason B.~S., Willman B.,
	Aguirre J.~E., 2014, ApJ, 795, L5
\bibitem[{Springel et al.}(2001)]{Springel2001} Springel V., Yoshida N., White S. D. M., 2001, New Astronomy, 6, 79
\bibitem[{Springel}(2005)]{Springel2005} Springel V., 2005, MNRAS, 364, 1105
\bibitem[van den Bergh(1972)]{VdB1972} van den Bergh S., 1972, ApJ, 171, L31
\bibitem[Walsh et al.(2007)]{Walsh2007} Walsh S.~M., Jerjen H., Willman B., 2007, ApJ, 662, L83
\bibitem[Wang et al.(2012)]{Wang2012} Wang J., Hammer F., Athanassoula E., Puech M., Yang Y., Flores H.,
	2012, A\&A, 538, A121
\bibitem[Watkins et al.(2010)]{Watkins2010} Watkins L.~L., Evans N.~W., An J.~H., 2010, MNRAS, 406, 264
\bibitem[Weisz et al.(2014)]{Weisz2014} Weisz D.~R., Dolphin A.~E., Skillman E.~D., Holtzman J.,
	Gilbert K. M., Dalcanton J. J., Williams B. F., 2014, ApJ, 789, 147
\bibitem[Widrow et al.(2003)]{Widrow2003} Widrow L.~M., Perrett K.~M., Suyu S.~H., 2003, \apj, 588, 311
\bibitem[{Widrow \& Dubinski}(2005)]{Widrow2005} Widrow L. M., Dubinski J., 2005, ApJ, 631, 838
\bibitem[{Widrow et al.}(2008)]{Widrow2008} Widrow L. M., Pym B., Dubinski J., 2008, ApJ, 679, 1239
\bibitem[Willman et al.(2005)]{Willman2005} Willman B. et al., 2005, ApJ, 626, L85
\bibitem[Young \& Lo(1997)]{Young1997} Young L.~M., Lo K.~Y., 1997, \apj, 476, 127
\bibitem[Zucker et al.(2006)]{Zucker2006} Zucker D.~B. et al., 2006, ApJ, 643, L103

\end{thebibliography}
\end{document}